\documentclass[12pt, draftclsnofoot, journal, onecolumn]{IEEEtran}
\usepackage{algorithm} 
\usepackage{algorithmic} 
\usepackage{multirow} 
\usepackage{amsmath}
\usepackage{xcolor}
\usepackage{amsmath}
\usepackage{stfloats}
\usepackage{amssymb}
\usepackage{amsmath}
\usepackage{cite}
\usepackage{url}
\usepackage{xcolor}
\usepackage{cite,graphicx,amsmath,amssymb}
\usepackage{subfigure}
\usepackage{citesort}
\usepackage{fancyhdr}
\usepackage{mdwmath}
\usepackage{mdwtab}
\usepackage{caption}
\usepackage{amsthm}
\usepackage{enumitem}
\usepackage{setspace}
\setenumerate{fullwidth,itemindent=\parindent,listparindent=\parindent,itemsep=0ex,partopsep=0pt,parsep=0ex}

\newtheorem{remark}{Remark}
\newtheorem{theorem}{Theorem}

\newtheorem{lemma}{Lemma}

\newtheorem{corollary}{Corollary}

\makeatletter
\def\ScaleIfNeeded{%
\ifdim\Gin@nat@width>\linewidth \linewidth \else \Gin@nat@width
\fi } \makeatother

\makeatletter
\renewcommand{\maketag@@@}[1]{\hbox{\m@th\normalsize\normalfont#1}}%
\makeatother

\begin{document}

\title{NOMA for STAR-RIS Assisted UAV Networks}

\author{
Jiayi~Lei,
Tiankui~Zhang,~\IEEEmembership{Senior Member,~IEEE,}
Xidong~Mu,
Yuanwei~Liu,~\IEEEmembership{Senior Member,~IEEE}

\thanks{
This work was supported by the National Key Research and Deployment Program of China under Grant 2019YFC1511302.
}

\thanks{Jiayi~Lei and Tiankui~Zhang are with the School of Information and Communication Engineering, Beijing University of Posts and Telecommunications, Beijing 100876, China (e-mail: \{leijiayi,zhangtiankui\}@bupt.edu.cn).}
\thanks{Xidong~Mu and Yuanwei~Liu are with the School of Electronic Engineering and Computer Science, Queen Mary University of London, London E1 4NS, U.K. (e-mail: \{x.mu, yuanwei.liu\}@qmul.ac.uk).}
}

\maketitle

\begin{abstract}
This paper proposes a novel simultaneously transmitting and reflecting reconfigurable intelligent surface (STAR-RIS) assisted unmanned aerial vehicle (UAV) non-orthogonal multiple access (NOMA) emergency communication network. Multiple STAR-RISs are deployed to provide additional and intelligent transmission links between trapped users and UAV-mounted base station (BS). Each user selects the nearest STAR-RIS for uploading data, and NOMA is employed for users located at the same side of the same STAR-RIS. Considering piratical requirements of post-disaster emergency communications, we formulate a throughput maximization problem subject to constraints on minimum average rate and maximum energy consumption, where the UAV trajectory, STAR-RIS passive beamforming, and time and power allocation are jointly optimized.
Furthermore, we propose a Lagrange based reward constrained proximal policy optimization (LRCPPO) algorithm, which provides an adaptive method for solving the long-term optimization problem with cumulative constraints. Specifically, using Lagrange relaxation, the original problem is transformed into an unconstrained problem with a two-layer structure. The inner layer is solved by penalized reward based proximal policy optimization (PPO) algorithm. In the outer layer, Lagrange multipliers are updated by gradient descent.
Numerical results show the proposed algorithm can effectively improve network performance while satisfying the constraints well. It also demonstrates the superiority of the proposed STAR-RIS assisted UAV NOMA network architecture over the benchmark schemes employing reflecting-only RISs and orthogonal multiple access.

\end{abstract}

\providecommand{\keywords}[1]{\textbf{\textit{Index terms---}}#1}
\begin{IEEEkeywords}
Emergency communication, resource allocation, simultaneously transmitting and reflecting reconfigurable intelligent surface, unmanned aerial vehicle.

\end{IEEEkeywords}

\newpage
\section{Introduction}
Large-scale natural disasters, such as earthquakes and hurricanes often inflict serious and unpredictable losses of life and property. The communication network plays a vital role in post-disaster rescue and recovery. However, conventional terrestrial communication infrastructures may be severely damaged during a disaster \cite{disaster_example}. For example, over 85\% of cell towers in affected area were inoperative during the Hurricane Harvey in the United States in 2017. In such scenarios, a temporal emergency communication network needs to be quickly established to restore connectivity and provide fast responses to emergency requests \cite{ECN1}. Considering the poor geographical circumstance and extreme urgency of post-disaster rescue missions, the emergency communication network has to be easily deployed, low-cost, and high-capacity.

Due to the characteristics of flexible deployment, affordable cost, and higher probability of line-of-sight (LoS) communication, the unmanned aerial vehicle (UAV)-based communication network has been one of the most potentially efficient candidates for emergency communications\cite{UAV1}. Nevertheless, there are still many technical bottlenecks for deploying UAVs as flying base stations (BSs) in disaster areas, including limited radio resource, adaptive trajectory design, and constrained UAV battery capacity. The insufficient battery capacity is the most intractable issue among above challenges, which has a significant negative impact on the energy efficiency and coverage of UAV, especially in the case of wide post-disaster areas \cite{UAV3}. As a result, there exists an urgent need for employing new technologies to make up for the drawbacks of UAV and fully unleash the potentials of UAV-based communications \cite{irs+uav1}.

As a promising technology, reconfigurable intelligent surfaces (RISs), or named intelligent reflecting surfaces (IRSs), have received significant attention \cite{9261597,9530750,9139273}. They are capable of significantly improving the performance of wireless communication networks by smartly reconfiguring the wireless propagation channel between the transmitter and the receiver. Conventional RISs mainly focus on reflecting the incident signal, so the transmitter and the receiver have to be located at the same side of the RIS. As a result, to face all users, conventional reflecting-only RISs must be deployed at the edge of the region of interest, which greatly limits the flexibility and efficiency of RISs. To address this issue, a novel simultaneously transmitting and reflecting RIS (STAR-RIS) is proposed in \cite{magazine_star}, which is also named intelligent omni-surface (IOS) in some works \cite{IOS}. Unlike reflecting-only RISs, the incident wireless signals can be transmitted and reflected by the STAR-RIS simultaneously. With this important characteristic, enhanced degrees of freedom (DoFs) for signal propagation manipulation are provided. STAR-RIS can serve all users at any location and full-space smart radio environment can be created. It has been confirmed in \cite{STAR+NOMA1} that the coverage of a STAR-RIS is 1.5 times as large as that of a reflecting-only RIS.  Therefore, it is more efficient to employ STAR-RISs instead of reflecting RISs in wireless communication networks.
Furthermore, employing the non-orthogonal multiple access (NOMA) technology in RIS or STAR-RIS assisted networks is a notable research topic. On the one hand, NOMA allows multiple users to simultaneously occupy the same time-and-frequency resource \cite{NOMA}, thereby achieving user fairness, massive connectivity, and spectral efficiency enhancement in RIS assisted network. On the other hand, by increasing the design flexibility and smartly changing the transmission channel, RISs especially STAR-RISs introduce important benefits in NOMA networks. The win-win effect of combining STAR-RIS and NOMA has been shown in several studies \cite{STAR+NOMA1,STAR+Secure}.

Thanks to the advantages of easy deployment, low cost, low power consumption, and $360^\circ$ smart radio environment, it is a feasible, economical, and effective scheme to employ STAR-RISs in UAV-based emergency communication networks. On the one hand, deploying STAR-RISs on-site by professionals after disasters is realizable for following three reasons. Firstly, compared with traditional communication devices that require computer rooms, STAR-RIS requires almost no additional equipment besides its own panel, making the deployment simple. Secondly, STAR-RIS is small in size, light in weight, and can be mounted on the surface of buildings, making the addressing easy. Thirdly, STAR-RIS can be composed of small panels, which is easy to expand and flexible to deploy. For disaster areas inaccessible to emergency personnel, deploying STAR-RISs would be more difficult. However, there are still solutions, such as employing unmanned emergency vehicles equipped with STAR-RISs, manipulating UAVs to drop STAR-RISs at designated positions and so on. On the other hand, employing STAR-RISs in UAV emergency networks is economical and efficient. First of all, as mentioned above, the hardware cost and deployment cost of STAR-RISs are both low. In addition, the power consumption of STAR-RIS is negligible, making it extremely suitable for post-disaster scenarios where energy is scarce. What's more, by creating full-space smart radio environment, STAR-RISs can make up for the short endurance and small coverage of single UAV, thereby improving the network performance. It is worth mentioning that deploying multiple UAV base stations is one of means for emergency communications. However, without doubt, it brings higher cost and energy consumption, and more complex management and design than deploying multiple STAR-RIS.

Driven by the aforementioned issues, this paper aims to investigate the STAR-RIS assisted UAV emergency networks with NOMA. In post-disaster scenarios, there are mainly three challenges for the STAR-RIS assisted UAV emergency networks. First of all, with the harsh electromagnetic environment and scarce radio resource after disasters, the UAV trajectory design and the radio resource allocation are very important for system performance improvement. Secondly, in order to make full use of the STAR-RISs, the passive beamforming design and STAR-RIS deployment are key topics that deserve investigation. Last but not least, compared with general communication scenarios, there are more stringent constraints and requirements in emergency communications. On the one hand, because of the paralysis of the ground power system after disasters \cite{UE_limited_energy}, the available energy for the trapped user equipment (UE) is severely limited, which brings additional challenge for efficient communications. On the other hand, to ensure post-disaster rescue for each user, the emergency communications have higher requirement on the minimum user transmission rate. In summary, it is a challenging but meaningful research topic to improve communication efficiency while satisfying the constraints on both UE rate and UE energy in STAR-RISs assisted UAV emergency communication networks, where UAV trajectory, STAR-RISs beamforming, and resource scheduling have to be carefully designed.

\vspace{-0.4cm}
\subsection{Related Works}
\vspace{-0.2cm}
\begin{enumerate}
  \item \emph{UAV-based Communication Networks:} During past years, UAVs have been applied in various communication scenarios and play different roles \cite{UAV+relay,UAV+MEC,UAV+SWIPT}. UAVs in \cite{UAV+relay} acted as mobile relay nodes for throughput improvement. In \cite{UAV+MEC}, a UAV equipped with an edge computing server was deployed to cope with computation-intensive tasks. In \cite{UAV+SWIPT}, the UAV was used to charge the ground sensor users and transmit information simultaneously.
      There is also some research focusing on the application of UAV in post-disaster emergency communications \cite{UAVEWC1,UAVEWC2,UAVEWC3}. Three UAV-assisted emergency communication schemes in different disaster scenarios were proposed in \cite{UAVEWC1}. Specifically, with surviving ground base stations, the flight trajectory and communication scheduling of UAVs were jointly optimized. In the scenarios without ground BSs, a UAV assisted Device-to-Device (D2D) communication system was studied. For the information exchange between disaster areas and outside, the hovering positions of UAV relays were optimized. In \cite{UAVEWC2}, in order to maximize the spectrum efficiency of the UAV-assisted emergency networks, the MBS power allocation, UAV zone selection and user scheduling were jointly optimized by a Deep-Q-Network-based algorithm.  An adaptive UAV deployment scheme was proposed in \cite{UAVEWC3} to achieve the maximal coverage for ground nodes after disasters. Most of these existing works related to UAV-based emergency communications focused on the UAV energy consumption, while the limited energy of the trapper users after disasters was ignored.
  \item \emph{(STAR-)RIS-Assisted Wireless Networks:} Recently, RIS-assisted wireless communication networks have been extensively studied \cite{RIS+channel_esitimation,RIS+MIMO,RIS+MEC}. In \cite{RIS+channel_esitimation}, the channel estimation for the RIS-enhanced orthogonal frequency division multiplexing (OFDM) system was investigated. A RIS-aided massive MIMO system was studied in \cite{RIS+MIMO}, where the active beamforming at the multi-antenna access point and the passive reflecting beamforming at the RIS were jointly optimized. In \cite{RIS+MEC}, the RIS was deployed in a mobile edge computing (MEC) system to reduce the average computation latency. As the incident signals can only be reflected by conventional RISs, in these RIS assisted networks, all users were located at one side of the RIS and access point (or base station) was assumed at the same side with users. To break through this assumption, increasing attention is paid to STAR-RIS, and some studies on the STAR-RIS assisted communication networks have been carried out \cite{STAR2,STAR+MIMO}. The authors of \cite{STAR2} proposed three practical operating protocols for the STAR-RIS, namely energy splitting, mode switching, and time switching. Then, a BS power consumption minimization problem in the STAR-RIS assisted downlink communications was
      considered for each protocol. In \cite{STAR+MIMO}, a STAR-RIS-assisted MIMO system was studied based on the energy splitting scheme. The precoding matrices and the transmitting and reflecting coefficients were optimized by the block coordinate descent algorithm.  What's more, owing to the unique dual-mode feature of the STAR-RIS, NOMA is naturally applied in STAR-RIS assisted networks \cite{STAR+NOMA1,STAR+Secure}. In \cite{STAR+NOMA1}, one transmitted user and one reflected user were paired as a NOMA cluster with the aid of the STAR-RIS. In \cite{STAR+Secure}, the STAR-RIS was combined with the NOMA technology, and the sum secrecy rate of the artificial noise aided communication network was maximized.
\item \emph{RIS-Assisted UAV Networks:} Currently, there have been a few works employing conventional reflecting-only RIS in UAV-based communication networks. In \cite{UAV+RIS5,uav_ris}, RIS was deployed on the ground and acted as a passive relay, while in \cite{UAV-RIS1,UAV-RIS3,UAV2}, RIS was mounted on the UAV and acted as an aerial mobile relay. Specifically, in \cite{UAV+RIS5}, RIS was involved to tackle the blocked link from UAV-BS (UBS) to ground terminals, by providing additional reflecting communication link. The 3D trajectory of the UAV and the phase shift of the RIS were designed to maximize the data transmission rate. In \cite{uav_ris}, the RIS was deployed to enhance the communication quality and reduce the movement of the UAV, thus saving the UAV energy. It is notable that in the above existing RIS assisted UAV networks, all users were located at one side of the RIS, and UAV also flew in the same side. In \cite{UAV-RIS1}, the author aimed to maximize the throughput of a RIS-UAV relaying communication system. In \cite{UAV-RIS3}, a UAV equipped with RIS was used to assist ground base station to cover users in hotspot areas. In \cite {UAV2}, multiple UAVs were equipped with the RISs and acted as aerial passive relays between the on-site command center and the trapped users in disaster areas. The author mainly focused on the composite fading channel and the joint bandwidth-power allocation, where the design of RIS beamforming and UAV trajectory were not involved.
\end{enumerate}

\vspace{-0.4cm}
\subsection{Motivations and Contributions}
\vspace{-0.2cm}
As summarized above, most of related works considered conventional reflecting-only RIS assisted UAV networks. As the incident signals can only be reflected by one side of RISs, there exist geographic constraints on transmitters and receivers, meaning that the UAV has to be on the same side of the RIS with users. In contrast, the incident signals can be reflected and transmitted by STAR-RIS simultaneously, thereby breaking through the geographic constraints, and fully reaping the benefits of UAVs and RISs. However, currently, few studies investigate the interplay between STAR-RIS and UAV communications [31,32]. In [31], the sum rate of users in STAR-RIS assisted UAV communication system was maximized, by jointly optimizing STAR-RIS beamforming, UAV trajectory and power allocation. In [32], the authors also aimed to improve the sum-rate by employing STAR-RIS in UAV communications, where the UAV was equipped with multiple antennas.
It is worth noting that these existing works involved single STAR-RIS, and employed conventional orthogonal multiple access (OMA) method, which didn't give full play to the advantages of STAR-RIS. What's more, none of them focused on the post-disaster emergency scenarios, where limited user energy and guaranteed minimum transmission rate need to be considered together.

Motivated by the observation, we propose a multiple STAR-RISs assisted UAV emergency communication network with NOMA. Our objective is to further explore the benefits of combining STAR-RIS and NOMA in UAV networks, and provide a feasible and efficient network architecture for post-disaster emergency communications. Furthermore, to tackle the cumulative constraints on user energy and rate in post-disaster scenarios, we involve a novel joint optimization algorithm, which is a combination of Lagrange relaxation and proximal policy optimization. The proposed algorithm updates penalty coefficients adaptively, enabling it to achieve the optimization objective while satisfying the constraints well. The main contributions of this paper are summarized as follows.

\begin{itemize}
  \item We propose a STAR-RIS assisted UAV NOMA emergency communication network, where a single UAV acts as the aerial BS and multiple STAR-RISs are deployed on the ground. Each UE accesses the nearest STAR-RIS to perform data uploading, and NOMA is employed for UEs located at the same side of the same STAR-RIS. Targeting the practical requirements of post-disaster communications, we formulate a throughput maximization problem subject to the cumulative constraints on the minimum UE average rate and the maximum UE energy consumption, where the UAV trajectory, STAR-RIS beamforming, and time and power allocation are jointly optimized.
  \item We formulae the throughput maximization problem as a Constrained Markov Decision Process (CMDP), and propose a Lagrange based reward constrained proximal policy optimization (LRCPPO) algorithm to solve it. Applying the Lagrange relaxation, the CMDP is converted into an unconstrained problem with a two-layer structure. Given the Lagrange multipliers, the inner layer problem is then formulated as a Markov Decision Process (MDP) with penalized reward function, and solved by proximal policy optimization (PPO) algorithm. As for the outer layer problem, the Lagrange multipliers are updated by gradient descent. The proposed LRCPPO algorithm provides an adaptive and effective method for solving long-term optimization problems with cumulative constraints.
  \item We illustrate the performance of the proposed network architecture and the LRCPPO algorithm by simulation. Three cases with different rate and energy constraints are considered. Numerical results show that 1)~the combination of STAR-RISs and NOMA attains significant throughput gains over the conventional reflecting-only RISs and OMA for UAV emergency communications; 2)~the proposed LRCPPO algorithm satisfies the constraints well in all cases,  and the UAV trajectory can be adaptively adjusted according to the constraints; 3)~the proposed LRCPPO algorithm evidently achieves higher throughput than the common reward shaping based algorithm and other benchmarks.
\end{itemize}

The rest of this paper is organized as follows. Section \uppercase\expandafter{\romannumeral2} introduces the system model of the proposed multiple STAR-RISs assisted UAV NOMA emergency communication network and formulates a long-term throughput maximization problem. In Section \uppercase\expandafter{\romannumeral3}, a Lagrange based reward constrained proximal policy optimization algorithm is proposed. Section \uppercase\expandafter{\romannumeral4} provides numerical results. Finally, Section \uppercase\expandafter{\romannumeral5} concludes this paper.

\newpage
\section{System Model and Problem Formulation}
After disasters, the original terrestrial communication system is almost paralyzed. The UAV-mounted BS is a promising solution to rebuild communication connections for post-disaster areas quickly. However, the endurance and coverage of UAVs are usually limited. Deploying STAR-RISs to assist the communication between the ground UEs and the UBS will be effective in further improving the efficiency of future emergency communication networks.

This paper focuses on the uplink transmission in the STAR-RIS assisted UAV emergency communication network. The list of notations is illustrated in Table. \ref{tabel1}.

\begin{table}[!h]
\begin{spacing}{1.1}
	\caption{ List of Notations}
    \vspace{-0.5cm}
    \label{tabel1}
	\begin{center}
		\begin{tabular}{|p{3cm}<{\centering}|c|}
            \hline
            \textbf{Notation} & \textbf{Description}  \\       \hline
			$I$              & Number of STAR-RISs  \\         \hline
			$K$                & Number of UEs   \\            \hline
            $M$                & Number of elements of STAR-RIS   \\            \hline
			${{\mathbf{q}}_k}$            & Location of UE $k$    \\      \hline
			${{\mathbf{q}}_i}$             & Location of STAR-RIS $i$ \\  \hline
			$\tau$        & Length of one time slot  \\        \hline
			${{\mathbf{q}}_U}[n]$     & Location of UAV at time slot $n$\\  \hline
            ${\mathcal{K}_i}$  & Set of UEs served by the STAR-RIS $i$ \\  \hline
            ${\mathcal{K}}_i^L$        & Set of UEs  located in the left space of STAR-RIS $i$    \\ \hline
            ${\mathcal{K}}_i^R$    & Set of UEs located in the right space of STAR-RIS $i$ \\ \hline
            $K_i^L$  & Number of elements in ${\mathcal{K}}_i^L$ \\ \hline
            $K_i^R$  & Number of elements in ${\mathcal{K}}_i^R$ \\ \hline
            $\tau _i^L[n]$   & Duration of STAR-RIS $i$ working for $K_i^L$ \\ \hline
            $\tau _i^R[n]$  & Duration of STAR-RIS $i$ working for $K_i^R$ \\ \hline
            $BW$            & Available bandwidth \\ \hline
            ${{\bf{h}}_{k - i}}[n] \in {\mathbb{C}^{M \times 1}}$   & Channel vector between UE $k$ and STAR-RIS $i$  \\ \hline
            ${{\bf{h}}_{i - U}}[n] \in {\mathbb{C}^{M \times 1}}$   & Channel vector between  STAR-RIS $i$  and UBS  \\ \hline
            ${{\bf{h}}_{k - U}}[n] \in {\mathbb{C}^{1 \times 1}}$   & Channel vector between  UE $k$  and UBS  \\ \hline
            $\zeta [n]$  & Reflecting / Transmitting flag \\ \hline
            ${\bf{\Lambda }}_i^r$  & Reflection phase shift matrix of STAR-RIS $i$   \\ \hline
            ${\bf{\Lambda }}_i^t$ & Transmission phase shift matrix of STAR-RIS $i$   \\ \hline
            ${P_k}$   & Transmit power  of UE $k$   \\ \hline
            ${w_k}[n]$   & Decoding  order  of UE $k$   \\ \hline
            ${E_{\max }}$  & Upper limit of UE  available energy   \\ \hline
            ${r_{\min }}$  & Lower limit of UE average rate   \\ \hline
		\end{tabular}
	\end{center}
\vspace{-1.0cm}
\end{spacing}
\end{table}

\newpage

\subsection{Network Architecture}

\begin{figure}[htp]
  \vspace{-0.2cm}
  \centering
  \includegraphics[width=5in]{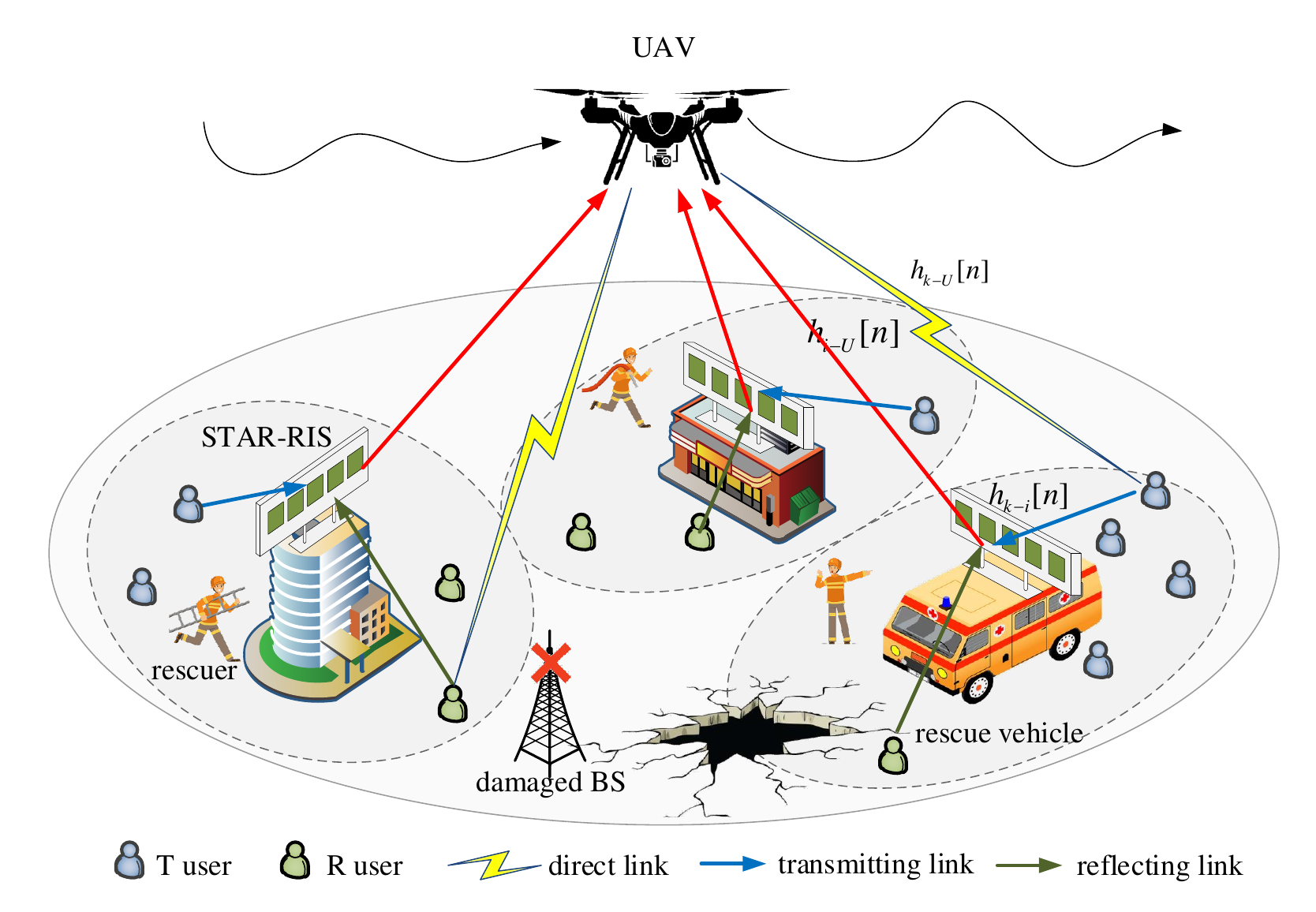}\\
  \caption{STAR-RIS assisted UAV emergency communication networks.}\label{Fig1}
  \vspace{-0.4cm}
\end{figure}
As shown in Fig. \ref{Fig1}, $K$ UEs are randomly distributed in a post-disaster area, where the ground BS is severely damaged. One UBS is dispatched and $I$ available STAR-RISs are deployed to recover the post-disaster communication. Both UBS and UE are equipped with single antenna. The STAR-RISs are operated with the time switching protocol, which means that STAR-RISs periodically switch all elements between transmitting mode and reflecting mode \cite{STAR2}.
It has been proved in \cite{9326394} that the optimal location of RIS should be close to the receiver or the transmitter. Inspired by this insightful conclusion, we assume that each UE selects the STAR-RIS which is closest to it to perform data uploading \footnote{In post-disaster scenarios, there are very likely to be only a few proper candidate locations for STAR-RISs. Based on this practical reality, we consider fixed and given locations of STAR-RISs in this paper.}.
The UE data signal received by the UBS is a superposition of the signals from two transmission links: direct link (UE--UBS link) and STAR-RIS aided link (UE--STAR-RIS--UBS link). The latter is further divided into transmitting link and reflecting link, depending on the relative locations of the UAV, UE and STAR-RIS. When the UAV and the UE are on the same side with respect to the corresponding STAR-RIS, the data uploading exploits the reflecting link via the STAR-RIS. Otherwise, the signal is uploaded to the UBS by exploiting the transmitting link.

Suppose that all STAR-RISs have the same number of elements, which is denoted by $M$. The 3D locations of UEs and STAR-RISs are assumed to be fixed, which are denoted by ${{\mathbf{q}}_k} = \left[ {{q_{k,x}},{q_{k,y}},{H_k}} \right]$ and ${{\mathbf{q}}_i} = \left[ {{q_{i,x}},{q_{i,y}},{H_i}} \right]$, respectively. The UAV flies at the fixed altitude of ${H_U}$ within the duration of $T$, and its maximum flight speed is ${v_{\max }}$. The duration $T$ is divided into $N$ time slots of the same length $\tau$. $\tau {v_{\max }} \ll {H_U}$, so the location of the UAV in each time slot is approximately regarded as unchanged. Then, the UAV flight trajectory is characterized by a set of vectors $ \mathbb{Q}= \left\{ {{{\mathbf{q}}_U}[n]} \right\}_{n = 1}^N$, where $\left\| {{{\mathbf{q}}_U}[n] - {{\mathbf{q}}_U}[n - 1]} \right\| \le {v_{\max }}\tau $.

\begin{figure}[htb]
  \centering
  \vspace{-1cm}
  \includegraphics[width=4.5in]{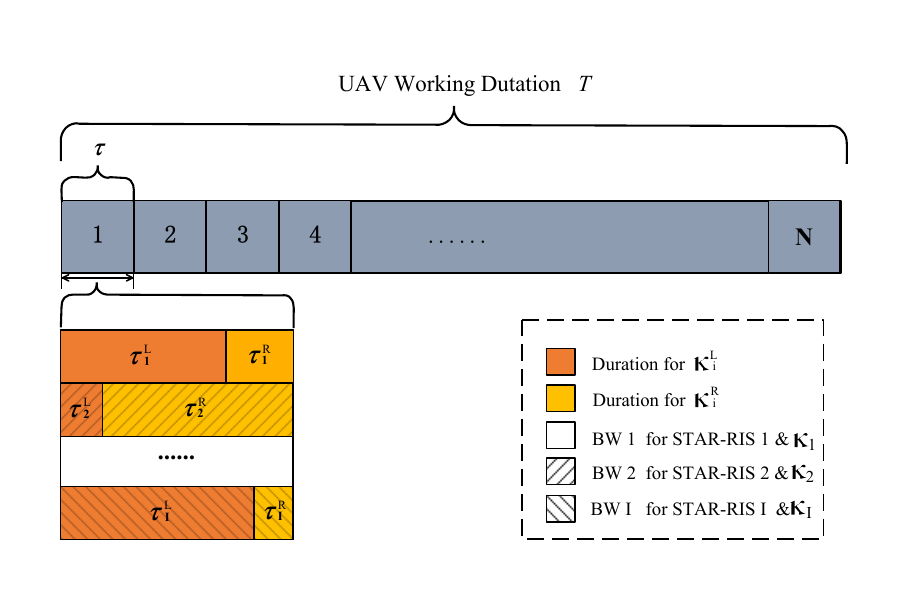}\\
  \caption{The framework of timescales and multiple access modes.}\label{Fig2}
\end{figure}

Let ${\mathcal{K}_i}$ denote the set of UEs served by the $i$-th STAR-RIS, where ${\mathcal{K}} = {\mathcal{K}_1} \cup {\mathcal{K}_2} \cup .... \cup {\mathcal{K}_I}$ and ${\mathcal{K}_i} \cap {\mathcal{K}_{i'}} = \emptyset ,\forall i,i' = 1,2,...,I$. For ${\mathcal{K}_i}$, the set of UEs located at the left side of the STAR-RIS $i$ is denoted by ${\mathcal{K}}_i^L = \{ k \in {\mathcal{K}_i}|{q_{k,x}} < {q_{i,x}}\} $, and the set of UEs located at the right side is denoted by ${\mathcal{K}}_i^R = \{ k \in {\mathcal{K}_i}|{q_{k,x}} \ge {q_{i,x}}\} $. The sizes of ${\mathcal{K}}_i^L$ and ${\mathcal{K}}_i^R$ are denoted by $K_i^L$ and $K_i^R$, respectively. Each STAR-RIS and the corresponding served UEs occupy orthogonal spectrum resources in the form of frequency division multiple access (FDMA). For the interior of ${{\mathcal{K}}_i}$, using the time division multiple access (TDMA), UEs in ${\mathcal{K}}_i^L$ and UEs in ${\mathcal{K}}_i^R$ are served sequentially. Then, UEs in ${\mathcal{K}}_i^L$ (${\mathcal{K}}_i^R$) upload data at the same time and the same frequency band via non-orthogonal multiple access (NOMA). The available frequency bandwidth of the system is denoted by $BW$. The duration allocated to ${\mathcal{K}}_i^L$ at time slot $n$ is denoted by ${\tau _i^L[n]}$ and the duration allocated to ${\mathcal{K}}_i^R$ is denoted by ${\tau _i^R[n]}$, where $\tau _i^L[n] + \tau _i^R[n] = \tau $. Note that ${\tau _i^L[n]}$ and ${\tau _i^R[n]}$ are also the time switching allocation that the STAR-RIS $i$ works in the transmitting mode or reflecting mode. The framework of timescales and multiple access modes is shown in Fig. \ref{Fig2}.

\vspace{-0.5cm}

\subsection{Channel Model}

At time slot $n$, the complex equivalent baseband channel vectors between the UE $k$ and the STAR-RIS $i$, between the STAR-RIS $i$ and the UBS, between the UE $k$ and the UBS are denoted by ${{\mathbf{h}}_{k - i}}[n] \in {\mathbb{C}^{M \times 1}}$, ${{\mathbf{h}}_{i - U}}[n] \in {\mathbb{C}^{M \times 1}}$, ${h_{k - U}}[n] \in {\mathbb{C}^{1 \times 1}}$, respectively. Both the large-scale fading and small-scale fading are captured, and modeled as Rician fading\cite{channel_model}. Specifically, the channel coefficient between the UE $k$ and the STAR-RIS $i$ is given by
\begin{equation}
	  \label{equation_2}
      {{{\mathbf{h}}_{k - i}}[n] = \sqrt {{\xi _{k - i}}} \left( {\sqrt {\frac{{{G_{k - i}}}}{{{G_{k - i}} + 1}}} {\mathbf{h}}_{k - i}^{LoS} + \sqrt {\frac{1}{{{G_{k - i}} + 1}}} {\mathbf{h}}_{k - i}^{NLoS}[n]} \right)},
      \end{equation}
where ${G_{k - i}}$ denotes the Rician factor. ${\xi _{k - i}}$ denotes the large-scale fading coefficient, which is determined by the distance ${d_{k - i}} = \left\| {{{\mathbf{q}}_k} - {{\mathbf{q}}_i}} \right\|$ and given by
      \begin{equation}
        {{\xi _{k - i}} = \frac{{{\xi _0}}}{{d_{k - i}^{{\alpha _1}}}}}.
      \end{equation}
${\xi _0}$ is the channel power at the reference distance of 1 meter, and ${\alpha _1}$ is the path loss exponent. ${\mathbf{h}}_{k - i}^{LoS}$ and ${\mathbf{h}}_{k - i}^{NLoS}[n]$ denote the line-of-sight (LoS) channel component and the non-line-of-sight (NLoS) component,  respectively. It is assumed that the STAR-RIS employs a uniform linear array (ULA) of elements, and then ${\mathbf{h}}_{k - i}^{LoS}$ is given by
      \begin{equation}
        {{\mathbf{h}}_{k - i}^{LoS} = {e^{ - j\frac{{2\pi {d_{k - i}}}}{\lambda }}} \times {\left[ {1,{e^{ - j\frac{{2\pi d}}{\lambda }\cos {\phi _{k - i}}}},...,{e^{ - j\frac{{2\pi (M - 1)d}}{\lambda }\cos {\phi _{k - i}}}}} \right]^T}},
      \end{equation}
where $d$ is the STAR-RIS element spacing, $\lambda$ denotes the carrier wavelength, and $\cos {\phi _{k - i}} = \frac{{{q_{i,x}} - {q_{k,x}}}}{{{d_{k - i}}}}$ is the cosine of the angle of arrival (AoA). The NLoS component of each element ${\mathbf{h}}_{k - i}^{NLoS}[n]$ is assumed to be independent and identically distributed and follow the circularly symmetric complex Gaussian distribution with zero mean and unit variance.

Similarly, the channel coefficient between the STAR-RIS $i$ and the UBS is
\begin{equation}
    {{{\mathbf{h}}_{i - U}}[n] = \sqrt {{\xi _{i - U}}[n]} \left( {\sqrt {\frac{{{G_{i - U}}}}{{{G_{i - U}} + 1}}} {\mathbf{h}}_{i - U}^{LoS}[n] + \sqrt {\frac{1}{{{G_{i - U}} + 1}}} {\mathbf{h}}_{i - U}^{NLoS}[n]} \right)},
\end{equation}
where ${\xi _{i - U}} = \frac{{{\xi _0}}}{{d_{i - U}^{{\alpha _2}}[n]}}$, ${d_{i - U}}[n] = \left\| {{\mathbf{q}_U}[n] - {\mathbf{q}_i}} \right\|$. ${\alpha _2}$ denotes the path loss exponent of the STAR-RIS--UBS link and ${G_{i - U}}$ denotes the Rician factor. The LoS component is given by
\begin{equation}
    {{\mathbf{h}}_{i - U}^{LoS}[n] = {e^{ - j\frac{{2\pi {d_{i - U}}[n]}}{\lambda }}} \times {\left[ {1,{e^{ - j\frac{{2\pi d}}{\lambda }\cos {\phi _{i - U}}[n]}},...,{e^{ - j\frac{{2\pi (M - 1)d}}{\lambda }\cos {\phi _{i - U}}[n]}}} \right]^T}},
\end{equation}
where $\cos {\phi _{i - U}}[n] = \frac{{{q_{U,x}}[n] - {q_{i,x}}}}{{{d_{i - U}}[n]}}$ is the cosine of the angle of departure (AoD). The NLoS component ${\mathbf{h}}_{i - U}^{NLoS}[n]$ also follows the circularly symmetric complex Gaussian distribution with zero mean and unit variance.

For the direct link (UE $k$--UBS), the channel vector is given as
\begin{equation}
    {{h_{k - U}}[n] = \sqrt {{\xi _{k - U}}[n]} \left( {\sqrt {\frac{{{G_{k - U}}}}{{{G_{k - U}} + 1}}} h_{k - U}^{LoS}[n] + \sqrt {\frac{1}{{{G_{k - U}} + 1}}} h_{k - U}^{NLoS}[n]} \right)},
\end{equation}
where ${\xi _{k - U}[n]} = \frac{{{\xi _0}}}{{d_{k - U}^{{\alpha _3}}[n]}}$, ${d_{k - U}[n]} = \left\| {{\mathbf{q}_U}[n] - {\mathbf{q}_k}} \right\|$. ${\alpha _3}$ denotes the path loss exponent of UE-UBS link and ${G_{k - U}}$ is the corresponding Rician factor. The LoS component is $h_{k - U}^{LoS}[n] = {e^{ - j\frac{{2\pi {d_{k - U}}[n]}}{\lambda }}}$ and the NLoS component follows $h_{k - U}^{NLoS}[n] \sim \mathcal{CN}(0,1)$.

According to the $i$-th STAR-RIS's location with respect to the UBS at time slot $n$, it can be determined whether ${\mathcal{K}}_i^L$ (${\mathcal{K}}_i^R$) is a reflected UE cluster or a transmitted UE cluster. $\zeta [n] = \{ 0,1\}$ is used to represent the reflection/transmission flag of UE clusters. The value is 1 for reflection and 0 for transmission.
\begin{equation}
	\label{equation_1}
   {\zeta ^{{\bf{\kappa }}_i^L}}[n] = \left\{ \begin{array}{l}
    1,{\kern 1pt} {\kern 1pt} {\kern 1pt} {\kern 1pt} {\kern 1pt} if{\kern 1pt} {\kern 1pt} {\kern 1pt} {q_{U,x}}[n]{\kern 1pt} {\kern 1pt} {\kern 1pt} \le {q_{i,x}}\\
    0,{\kern 1pt} {\kern 1pt} {\kern 1pt} {\kern 1pt} {\kern 1pt} {\kern 1pt} if{\kern 1pt} {\kern 1pt} {\kern 1pt} {q_{U,x}}[n]{\kern 1pt} {\kern 1pt} {\kern 1pt}  > {q_{i,x}},
\end{array} \right.
    {\zeta ^{{\bf{\kappa }}_i^R}}[n] = \left\{ \begin{array}{l}
    1,{\kern 1pt} {\kern 1pt} {\kern 1pt} {\kern 1pt} {\kern 1pt} if{\kern 1pt} {\kern 1pt} {\kern 1pt} {q_{U,x}}[n]{\kern 1pt} {\kern 1pt} {\kern 1pt}  > {q_{i,x}}\\
    0,{\kern 1pt} {\kern 1pt} {\kern 1pt} {\kern 1pt} {\kern 1pt} {\kern 1pt} if{\kern 1pt} {\kern 1pt} {\kern 1pt} {q_{U,x}}[n]{\kern 1pt} {\kern 1pt} {\kern 1pt}  \le {q_{i,x}}.
\end{array} \right.
\end{equation}

The transmission coefficient matrix and the reflection coefficient matrix of STAR-RIS $i$ at time slot $n$ are represented by the diagonal matrix ${\bf{\Lambda }}_i^t[n] = \mathrm{diag}\left( {{e^{i\theta _{i1}^t[n]}},{e^{i\theta _{i2}^t[n]}},...,{e^{i\theta _{iM}^t[n]}}} \right)$ and ${\bf{\Lambda }}_i^r[n] = \mathrm{diag}\left( {{e^{i\theta _{i1}^r[n]}},{e^{i\theta _{i2}^r[n]}},...,{e^{i\theta _{iM}^r[n]}}} \right)$ respectively. $\theta _{im}^t[n],\theta _{im}^r[n] \in [0,2\pi )$ denote the reflection phase shift and the transmission phase shift of the $m$-th element of STAR-RIS $i$. Finally, the superimposed channel vector for uplink transmission of UE $k$ at time slot $n$ is expressed as
\begin{equation}
\begin{small}
{{h_k}[n] = \left\{ \begin{array}{l}
\overbrace {{\bf{h}}_{i - U}^H[n]\left( {{\zeta ^{{\bf{\kappa }}_i^L}}[n]{\bf{\Lambda }}_i^r[n] + \left( {1 - {\zeta ^{{\bf{\kappa }}_i^L}}[n]} \right){\bf{\Lambda }}_i^t[n]} \right){{\bf{h}}_{k - i}}[n]}^{STAR - RIS{\kern 1pt} {\kern 1pt} {\kern 1pt} aided{\kern 1pt} {\kern 1pt} {\kern 1pt} {\kern 1pt} link} + \overbrace {{h_{k - U}}[n]}^{direct{\kern 1pt} {\kern 1pt} {\kern 1pt} link},{\kern 1pt} {\kern 1pt} {\kern 1pt} {\kern 1pt} {\kern 1pt} {\kern 1pt} {\kern 1pt} {\kern 1pt} for{\kern 1pt} {\kern 1pt} {\kern 1pt} k \in {\mathcal{K}}_i^L,i = 1,2,...,I\\
\overbrace {{\bf{h}}_{i - U}^H[n]\left( {{\zeta ^{{\bf{\kappa }}_i^R}}[n]{\bf{\Lambda }}_i^r[n] + \left( {1 - {\zeta ^{{\bf{\kappa }}_i^R}}[n]} \right){\bf{\Lambda }}_i^t[n]} \right){{\bf{h}}_{k - i}}[n]}^{STAR - RIS{\kern 1pt} {\kern 1pt} {\kern 1pt} aided{\kern 1pt} {\kern 1pt} {\kern 1pt} {\kern 1pt} link} + \overbrace {{h_{k - U}}[n]}^{direct{\kern 1pt} {\kern 1pt} {\kern 1pt} link},{\kern 1pt} {\kern 1pt} {\kern 1pt} {\kern 1pt} {\kern 1pt} for{\kern 1pt} {\kern 1pt} {\kern 1pt} k \in {\mathcal{K}}_i^R,i = 1,2,...,I{\kern 1pt}
\end{array} \right.}.
\end{small}
\end{equation}

\subsection{NOMA Communications}
The UBS receives the signal from UEs in ${\mathcal{K}}_i^L$ (${\mathcal{K}}_i^R$) using NOMA. Successive interference cancelation (SIC) is employed at the UBS, and multi-UEs-signals are decoded in the descending order of channel gain. The decoding order for the UE $k$ at time slot $n$ is denoted by ${\omega _k}[n]$ and expressed as
\begin{equation}
{{\omega _k}[n] = \left\{ \begin{array}{l}
\{ 1,2,...,K_i^L\} ,{\kern 1pt} {\kern 1pt} {\kern 1pt} {\kern 1pt} {\kern 1pt} {\kern 1pt} \forall{\kern 1pt} {\kern 1pt} {\kern 1pt} k \in {\mathcal{K}}_i^L,i = 1,2,...,I\\
\{ 1,2,...,K_i^R\} ,{\kern 1pt} {\kern 1pt} {\kern 1pt} {\kern 1pt} {\kern 1pt} \forall{\kern 1pt} {\kern 1pt} {\kern 1pt} k \in {\mathcal{K}}_i^R,i = 1,2,...,I
\end{array} \right..}
\end{equation}
The transmission power of the UE $k$ at time slot $n$ is represented by ${P_k[n]}$. Based on the above network architecture and channel model, the average transmission rate of UE $k$ at time slot $n$ is given by
\begin{equation}
{{r_k}[n] = \left\{ \begin{array}{l}
\frac{{\tau _i^L[n]}}{\tau  I }{\log _2}\left( {1 + \frac{{{p_k}[n]{{\left| {{h_k}[n]} \right|}^2}}}{{\sum\limits_{{\omega _{k'}}[n] \ge {\omega _k}[n],k' \in \mathcal{K} _i^L} {{p_{k'}}[n]{{\left| {{h_{k'}}[n]} \right|}^2}}  + {\sigma ^2}}}} \right),{\kern 1pt} {\kern 1pt} {\kern 1pt} {\kern 1pt} {\kern 1pt} \forall{\kern 1pt} {\kern 1pt} {\kern 1pt} k \in {\mathcal{K}}_i^L,i = 1,2,...,I\\
\frac{{\tau _i^R[n]}}{\tau I }{\log _2}\left( {1 + \frac{{{p_k}[n]{{\left| {{h_k}[n]} \right|}^2}}}{{\sum\limits_{{\omega _{k'}}[n] \ge {\omega _k}[n],k' \in \mathcal{K} _i^R} {{p_{k'}}[n]{{\left| {{h_{k'}}[n]} \right|}^2}}  + {\sigma ^2}}}} \right),{\kern 1pt} {\kern 1pt} {\kern 1pt} {\kern 1pt} {\kern 1pt} \forall{\kern 1pt} {\kern 1pt} {\kern 1pt} k \in {\mathcal{K}}_i^R,i = 1,2,...,I
\end{array} \right.,}
\end{equation}
where ${\sigma ^2}$ is the noise power at the UBS receiver.

\subsection{Problem Formulation}
In post-disaster emergency communication scenarios, a large throughput is desired for collecting UEs data. At the same time, practical constraints have to be considered. On the one hand, a minimum transmission rate guarantee is required to ensure a reliable communication link for each trapped user. On the other hand, due to the paralyzed power system after disasters, the energy storage of UEs would be limited and cannot be charged in time. It is assumed that the maximum available energy for each UE in the duration $T$ is ${E_{\max }}$. By jointly optimizing the UAV trajectory ${\mathbf{q}_U}[n]$, the UEs transmit power ${P_k}[n]$, the STAR-RIS time switching allocation ${\tau _i}[n] = \{ \tau _i^L[n],\tau _i^R[n]\}$, and the STAR-RIS phase shift matrix ${\bf{\Lambda }}_i^r[n],{\bf{\Lambda }}_i^t[n]$, our goal is to maximize the long-term uplink throughput of the proposed emergency communication network while satisfying constraints on the minimum required UE average rate and maximum UE available energy \footnote{The proposed network structure can be applied in other scenarios where the existing infrastructure fails to meet communication needs, such as large-scale sports events, but the network objective and constraint involved should be designed based on actual situation.}. The optimization problem is formulated as follows.

\begin{subequations}\label{P1}
\begin{align}
~~&\mathop{{\mathrm{Max}}}\limits_{{\mathbf{q}_U}[n],{P_k}[n],{{\bf{\tau }}_i}[n],{\bf{\Lambda }}_i^r[n],{\bf{\Lambda }}_i^t[n]} \frac{1}{{N}}\sum\limits_{n = 1}^N {\sum\limits_{k = 1}^K {{r_k}[n]} },\\
~~\text{s.t.}~~~&\frac{1}{N}\sum\limits_{n = 1}^N {{r_k}[n]} {\kern 1pt}  \ge {r_{\min }},{\kern 1pt} {\kern 1pt} {\kern 1pt} {\kern 1pt} {\kern 1pt} {\kern 1pt} {\kern 1pt} {\kern 1pt} {\kern 1pt} \forall k \in \mathcal{K} ,\label{a}\\
&\sum\limits_{n = 1}^N {{P_k}[n]} \tau _i^L[n] \le {E_{\max }},{\kern 1pt} {\kern 1pt} {\kern 1pt} {\kern 1pt} {\kern 1pt} \forall k \in \mathcal{K} _i^L,i = 1,2,...,I,\label{b}\\
&\sum\limits_{n = 1}^N {{P_k}[n]} \tau _i^R[n] \le {E_{\max }},{\kern 1pt} {\kern 1pt} {\kern 1pt} {\kern 1pt} \forall k \in \mathcal{K} _i^R,i = 1,2,...,I,\label{c}\\
&\left\| {{\mathbf{q}_U}[n] - {\mathbf{q}_U}[n - 1]} \right\| \le {v_{\max }}\tau ,{\kern 1pt} {\kern 1pt} {\kern 1pt} {\kern 1pt} {\kern 1pt} n = 1,2,....,N,\label{d}\\
&{\mathbf{q}_U}[0] = {\mathbf{q}_0},\label{e}\\
&{P_k}[n] \le {P_{\max }},{\kern 1pt} {\kern 1pt} {\kern 1pt} n = 1,2,....,N,k = n = 1,2,....,K,\label{f}\\
&\tau _i^L[n] + \tau _i^R[n] = \tau ,{\kern 1pt} {\kern 1pt} {\kern 1pt} i = 1,2,...,I,n = 1,2,....,N,\label{g}\\
&\theta _{im}^t[n],\theta _{im}^r[n] \in [0,2\pi ),m = 1,2,...,M,i = 1,2,...,I,n = 1,2,...,N.\label{h}
\end{align}
\end{subequations}

Constraint (\ref{a}) represents the  UE average rate could not be lower than ${r_{\min }}$. Constraints (\ref{b}) and (\ref{c}) represent the energy consumed by UE in the duration $T$ should be less than ${E_{\max }}$. Constraint (\ref{d}) indicates the maximal UAV flight speed is ${v_{\max }}$. Constraint (\ref{e}) means the UAV takes off at the fixed position ${\mathbf{q}_0}$ and (\ref{f}) represents the maximum transmit power of UE is ${P_{\max }}$.

To tackle this non-convex optimization problem, we mainly face following three challenges. Firstly, the four optimization variables are highly coupled with each other, so it is difficult and complicated to divide the original problem into sub-problems and then solve them successively. Secondly, as the formulated problem is long-term and dynamic, achieving long-term optimization by solving per-single-slot optimization is not rigorous in theory. Thirdly, unlike the constraints on a single time slot\cite{single_time_slot1,single_time_slot2}, the average rate and the energy consumption of (\ref{a}) and (\ref{b}),~(\ref{c}) are long-term and cumulative, bringing difficulties to the design of the reward function in Markov decision process. Therefore, traditional reinforcement learning methods are not suitable to be applied directly.

\section{ Lagrange Based Reward Constrained Proximal Policy Optimization Algorithm}

In this section, we propose a Lagrange based reward constrained proximal policy optimization algorithm, abbreviated as LRCPPO, to solve (\ref{P1}). First of all, the long-term and dynamic optimization problem is formulated as a Constrained Markov Decision Process (CMDP). Secondly, applying the Lagrange Relaxation method, the CMDP is converted into an unconstrained problem with a min-max two-layer structure. The maximization problem in the inner layer is formulated as a Markov Decision Process with the penalized reward function, and solved by the reward constrained proximal policy optimization algorithm. The minimization problem in the outer layer is solved by gradient descent. The proposed LRCPPO algorithm updates relevant parameters on three timescales, and converges to a feasible suboptimal solution.
\subsection{CMDP}
A fundamental CMDP is defined by the tuple $\left\{ {S,A,r,P,{s_0}, c,\delta} \right\}$ \cite{CMDP}. $S$ denotes the state space. $A$ denotes the action space. $r:S \times A \times S \rightarrow \mathbb{R}$ represents the reward function. $P:S \times A \times S \rightarrow [0,1]$ is the state transition probability matrix and ${s_0}$ is the initial state. $c:S \times A \times S \rightarrow \mathbb{R} $ denotes the immediate constraint function (also called penalty function) at one time slot. $\delta \in [0,1]$ is the threshold of the corresponding long-term cumulative constraint function. In our proposed system model and optimization problem, the above seven elements are defined as follows.

\begin{itemize}
  \item  State $S$. The state in time slot $n$ is the UAV 2D location $\left[ {{q_{U,x}}[n],{q_{U,y}}[n]} \right]$.
  \item  Action	$A$. The action is composed of the UAV flight direction (denoted by the angle ${\theta _U}[n] \in [0,2\pi )$), the UAV flight speed $v[n] \in [0,{v_{\max }}]$, the UE transmit power ${P_k}[n]$, the time switching allocation ${{\bf{\tau }}_i}[n] = \{ \tau _i^L[n],\tau _i^R[n]\} $ and the beamforming matrix ${\bf{\Lambda }}_i^r[n], {\bf{\Lambda }}_i^t[n]$.
  \item  	Reward function $r$. The reward function at time slot $n$ is defined as the sum of the uplink rates of all UEs, $r({s_n},{a_n}) = \sum\limits_{k = 1}^K {{r_k}[n]}$.
  \item  	State transition probability matrix $P$. The UAV position at next time slot is fully determined by the position and the action at the current time slot, so each element of $P$ takes the value of $1$.
  \item  	Initial state ${s_0}$ is the UAV starting point ${\mathbf{q}_0} = [0,0,{H_U}]$.
  \item Constraint function $c$. For the rate constraint in (\ref{P1}), the immediate rate constraint at time slot $n$ is defined as
      \begin{equation}
       {c_{r,k}}({s_n},{a_n}) = {r_{\min }} - {r_k}[n],k \in \mathcal{K}.
      \end{equation}
      For the energy constraint in (\ref{P1}), the immediate energy constraint at time slot $n$ is
      \begin{equation}
       {{c_{e,k}}({s_n},{a_n}) = \left\{ \begin{array}{l}
       {P_k}[n]\tau _i^L[n] - \frac{{{E_{\max }}}}{N},\forall k \in \mathcal{K} _i^L,i = 1,2,...,I\\
       {P_k}[n]\tau _i^R[n] - \frac{{{E_{\max }}}}{N},\forall k \in \mathcal{K}_i^R,i = 1,2,...,I
      \end{array} \right.}.
      \end{equation}
 \item Constraint threshold $\delta$. Based on the definition of constraint functions, the constraint (\ref{a}) is equivalently expressed as $\sum\limits_{n = 1}^N {{c_{r,k}}({s_n},{a_n})} \le 0 ,k \in \mathcal{K}$, and the constraints (\ref{b}) and (\ref{c}) are equivalent to ${\sum\limits_{n = 1}^N {{c_{e,k}}({s_n},{a_n})} \le 0,k \in \mathcal{K}}$. Therefore, $ {\delta _{r,k}} = {\delta _{e,k}} = 0$ in this paper.
\end{itemize}

Policy $\pi :S \to {\Delta _A}$ is defined as the probability distribution over actions and $\pi (a|s) \in [0,1]$ denotes the probability of taking action $a$ at state $s$. The set of all possible policies is denoted by $\Pi$ and the objective function guided by $\pi$ is given by
\begin{equation}
    {J_R^\pi= {\mathbb{E}^\pi }[\sum\limits_n {r({s_n},{a_n})} |{s_0} = {\mathbf{q}_0}]}.
\end{equation}
Similarly, the cumulative rate constraint function can be expressed as
\begin{equation}
{J_{C_{r,k}}^\pi= {\mathbb{E}^\pi }[\sum\limits_n {c_{r,k}({s_n},{a_n})} |{s_0} = {\mathbf{q}_0}]},
\end{equation}
and the cumulative energy constraint function is given by
\begin{equation}
{J_{C_{e,k}}^\pi= {\mathbb{E}^\pi }[\sum\limits_n {c_{e,k}({s_n},{a_n})} |{s_0} = {\mathbf{q}_0}]}.
\end{equation}

Then, the optimization problem (\ref{P1}) can be equivalently expressed as
\begin{subequations}\label{P2}
\begin{align}
   ~~~~ & \mathop {\max }\limits_{\pi  \in \Pi } J_R^\pi, \\
   ~~\text{s.t.}~~~ & J_{{C_r}_{,k}}^\pi  \le 0,k \in \mathcal{K}, \\
   & J_{{C_e}_{,k}}^\pi  \le 0 {\kern 1pt} ,k \in \mathcal{K}, \\
   &(\ref{d})\sim(\ref{h}).
\end{align}
\vspace{-1.5cm}
\end{subequations}
\subsection{Lagrange Relaxation}
We employ the Lagrange relaxation method to solve the CMDP \cite{single_time_slot2}. Given (\ref{P2}), the Lagrange function is defined as
\begin{equation}\label{Lagrange_function}
  L({\lambda _{r,k}},{\lambda _{e,k}},\pi ){\rm{ = }}J_R^\pi  - \sum\limits_{k = 1}^K {{\lambda _{r,k}} {J_{{C_{r,k}}}^\pi} }  - \sum\limits_{k = 1}^K {{\lambda _{E,k}}{J_{{C_{e,k}}}^\pi}},
\end{equation}
where ${\lambda _{r,k}} \ge0$ is the Lagrange multiplier (or understood as a penalty coefficient) for the rate constraint of the UE $k$ and ${\lambda _{e,k}} \ge 0$ is the Lagrange multiplier for the energy constraint of the UE $k$. Then the optimization problem is converted into following unconstrained problem
\begin{equation}\label{P3}
~~~~\mathop {\min }\limits_{{\lambda _{r,k}} \ge {\rm{0}}{\lambda _{e,k}} \ge {\rm{0}}} \mathop {\max }\limits_\pi  L({\lambda _{r,k}},{\lambda _{e,k}},\pi ){\kern 1pt} {\kern 1pt} {\kern 1pt} {\kern 1pt} {\kern 1pt} {\kern 1pt} s.t.{\kern 1pt} {\kern 1pt} {\kern 1pt} {\kern 1pt} (\ref{d})\sim(\ref{h}).
\end{equation}

Due to the weak duality, there is a gap between the solution of (\ref{P2}) and that of the Lagrange dual problem (\ref{P3}). In fact, the solution of (\ref{P2}) provides an upper bound for the target solution. What's more, as the Lagrange multipliers increase, the solution to (\ref{P3}) would finally converge to that of (\ref{P2}) \cite{RCPO}.
Therefore, our goal is transformed into finding the saddle point $\left( {{\pi ^*}({\lambda _{r,k}}^*,{\lambda _{e,k}}^*),{\lambda _{r,k}}^*,{\lambda _{e,k}}^*} \right)$ of (\ref{P3}). As (\ref{P3}) is a two-layer optimization problem, we apply a two-step approach to solve it. For the inner layer, $\mathop {\max }\limits_\pi  L({\lambda _{r,k}},{\lambda _{e,k}},\pi )$, is solved with given ${\lambda _{r,k}},{\lambda _{e,k}}$. For the outer layer, the Lagrange multipliers are increased by gradient descent until corresponding constraints are satisfied.
\subsection{The inner layer : Reward Constrained Proximal Policy Optimization}
In this subsection, we propose a reward constrained proximal policy optimization method to solve the inner layer of (\ref{P3}), $\mathop {\max }\limits_\pi  L({\lambda _{r,k}},{\lambda _{e,k}},\pi )$, with given ${\lambda _{r,k}}$ and ${\lambda _{e,k}}$.

Employing the Lagrange relaxation, the constraints (\ref{a}) $\sim$ (\ref{c}) are incorporated into the initial objective function. As a result, the maximization problem $\mathop {\max }\limits_\pi  L({\lambda _{r,k}},{\lambda _{e,k}},\pi )$ in the inner layer can be formulated as a fundamental MDP $\left\{ {S,A,\hat r,P,{s_0}} \right\}$ with given Lagrange multipliers. The state, action, state transition probability matrix, and initial state of the MDP are completely the same with that of the CMDP. However, it is worth noting that the reward function of the MDP is redefined and named as penalized reward.

\textbf{Penalized Reward.}
The penalized reward function is defined as
\begin{equation}\label{penalized_reward}
    \hat r({\lambda _{r,k}},{\lambda _{e,k}}{s_n},{a_n}) = r({s_n},{a_n}) - \sum\limits_{k = 1}^K {{\lambda _{r,k}}{c_{r,k}}({s_{n,}}{a_n})}  - \sum\limits_{k = 1}^K {{\lambda _{e,k}}{c_{e,k}}({s_{n,}}{a_n})}.
\end{equation}
Note that the penalty coefficients in (\ref{penalized_reward}) are acted by the Lagrange multipliers.

\begin{remark}\label{remark1}
A reward shaping method is applied to tackle the constraints of the CMDP in some reinforcement learning algorithm based works \cite{uav_ris,reward_shaping2,reward_shaping3}. This method also adds penalty terms to the initial reward. But the penalty coefficients in the reward shaping are determined under manual selection, which brings two serious drawbacks: On the one hand, the manual selection of coefficients is a time-consuming and computationally intensive process of hyper-parameters tuning. It will be very tricky when the number of constraints increases. On the other hand, reward shaping is a non-adaptive approach. Once the communication network parameters change, the penalty coefficients need to be re-selected. The Lagrange based penalized reward proposed in this paper is able to make up for the drawbacks of the reward shaping and has higher interpretability.
\end{remark}

Then the corresponding penalized objective function is give by
\begin{small}
\begin{equation}\label{penalized_objective_function}
   {{\hat J}^\pi }\left( {{\lambda _{r,k}},{\lambda _{e,k}},s} \right)
    = {\mathbb{E}^\pi }\left[ {\left. {\sum\limits_n {\hat r\left( {{\lambda _{r,k}},{\lambda _{e,k}},{s_n},{a_n}} \right)} } \right|{s_0} = {\mathbf{q}_0}} \right],
\end{equation}
\end{small}
\noindent and the MDP is expressed as $\mathop {\max }\limits_\pi  {\hat J^\pi }\left( {{\lambda _{r,k}},{\lambda _{e,k}},s} \right) s.t.{\kern 1pt} {\kern 1pt} {\kern 1pt} {\kern 1pt} (\ref{d})\sim(\ref{h}).$ To solve this MDP, the proximal policy optimization (PPO) \cite{PPO} algorithm is used. PPO is an Actor-Critic based reinforcement learning algorithm, which is able to tackle MDP problems with continuous action efficiently. More importantly, by introducing the proximal optimization and clipping functions, the PPO has high data utilization and is easy to converge. Next, we illustrate how the PPO algorithm works in two parts: Critic network and Actor network.

\textbf{Critic Network.}
The input of the Critic neural network is the state, and the output is the corresponding estimated value. The parameters of the Critic network are denoted by ${\theta _c}$. At each iteration, $\left\{ {{s_j}} \right\}_{j = 1}^{|B|}$ in a certain number of samples $B = \left\{ {{s_j},{a_j},{{\hat r}_j},{{s'}_j}} \right\}_{j = 1}^{|B|}$ is input, and $\left\{ {{V^{{\theta _c}}}({s_j},{\lambda _{r,k}},{\lambda _{E,k}})} \right\}_{j = 1}^{|B|}$ is obtained. $|B|$ is the batch size of sample memory. Then the target value function is calculated by
\begin{equation}\label{value}
\begin{split}
  {V'^{{\theta _c}}}(s,{\lambda _{r,k}},{\lambda _{e,k}})  & =\sum\limits_n {{\gamma ^n}\hat r({\lambda _{r,k}},{\lambda _{e,k}},{s_n},{a_n})} |{s_0} = s \\
    & = \hat r({\lambda _{r,k}},{\lambda _{e,k}},s,a) + \gamma {V'^{{\theta _c}}}(s',a',{\lambda _{r,k}},{\lambda _{e,k}})|s,
\end{split}
\end{equation}
where $\gamma  \in [0,1]$ is the discount factor for future penalized rewards and ${V'^{{\theta _c}}}({s'_{|B|}},{\lambda _{r,k}},{\lambda _{e,k}}) = {V^{{\theta _c}}}({s'_{|B|}},{\lambda _{r,k}},{\lambda _{e,k}})$. The temporal difference (TD) error (loss function) of the Critic network is defined as
\begin{equation}\label{td_error}
    Loss({\theta _c},{\lambda _{r,k}},{\lambda _{e,k}}) = \frac{1}{{|B|}}{\sum\limits_{j = 1}^{|B|} {\left( {{{V'}^{{\theta _c}}}({s_j},{\lambda _{r,k}},{\lambda _{e,k}}) - {V^{{\theta _c}}}(s,{\lambda _{r,k}},{\lambda _{e,k}})} \right)} ^2}.
\end{equation}
${A^{{\theta _c}}}({s_j},{\lambda _{r,k}},{\lambda _{E,k}}) = {V'^{{\theta _c}}}({s_j},{\lambda _{r,k}},{\lambda _{E,k}}) - {V^{{\theta _c}}}({s_j},{\lambda _{r,k}},{\lambda _{E,k}})$ is called the advantage function at the state ${s_j}$. The critic network is trained by minimizing the TD-error, that is
\begin{equation}\label{critic_update}
    {\theta _c}^\prime  = {\theta _c} - {\eta _c}{\nabla _{{\theta _c}}}Loss({\theta _c}),
\end{equation}
where ${\eta _c}$ is the updating step-size (learning-rate) for $\theta _c$.

\textbf{Actor Network.} The input of the Actor network is also the state, and the output is the parameterized policy. To make more effective use of data, the importance sampling technology is applied in the Actor network of PPO. Specifically, the Actor network consists of two policy neural networks with completely the same structure: the behavior policy network (for generating data) and the target policy network (for interacting with environment). The parameters are denoted by ${\theta _{a\_behavior}}$ and ${\theta _{a\_target}}$ respectively. Note that only the target policy network ${\theta _{a\_target}}$ is trained, while the ${\theta _{a\_behavior}}$ is merely copied from ${\theta _{a\_target}}$ at intervals of fixed iterations.
At each iteration, $\left\{ {{s_j}} \right\}_{j = 1}^{|B|}$ is fed into the two policy networks, and then the probabilities of $\left\{ {{a_j}} \right\}_{j = 1}^{|B|}$ are obtained, denoted as $\left\{ {{p_{{\theta _{a\_target}}}}\left( {{a_j}|{s_j}} \right)} \right\}_{j = 1}^{|B|}$ and $\left\{ {{p_{{\theta _{a\_behavior}}}}\left( {{a_j}|{s_j}} \right)} \right\}_{j = 1}^{|B|}$ respectively. The importance weight of the action ${{a_j}}$ is defined as
\begin{equation}\label{IW}
    IW({a_j}) = \min \left\{ {\frac{{{p_{{\theta _{a\_target}}}}\left( {{a_j}|{s_j}} \right)}}{{{p_{{\theta _{a\_behavior}}}}\left( {{a_j}|{s_j}} \right)}},clip\left( {\frac{{{p_{{\theta _{a\_target}}}}\left( {{a_j}|{s_j}} \right)}}{{{p_{{\theta _{a\_behavior}}}}\left( {{a_j}|{s_j}} \right)}},1 - \varepsilon ,1 + \varepsilon } \right)} \right\},
\end{equation}
where $\varepsilon$ is a hyper-parameter. Then the loss function of the Actor network is given by
\begin{equation}\label{A_loss}
    Loss({\theta _{a\_target}}) = \frac{1}{{|B|}}\sum\limits_{j = 1}^{|B|} {\left( {\mathop {IW}({a_j}){A^{{\theta _c}}}\left( {{s_j},{\lambda _{r,k}},{\lambda _{e,k}}} \right)} \right)}.
\end{equation}
The network parameters are updated by
\begin{equation}\label{actor_update}
   {\theta '_{a\_target}} = {\theta _{a\_target}} + {\eta _a}{\nabla _{{\theta _{a\_target}}}}Loss({\theta _{a\_target}}),
\end{equation}
where ${\eta _a}$ is the learning rate for the Actor.

\begin{figure}[hptb]
  \centering
  \includegraphics[width=4.2in]{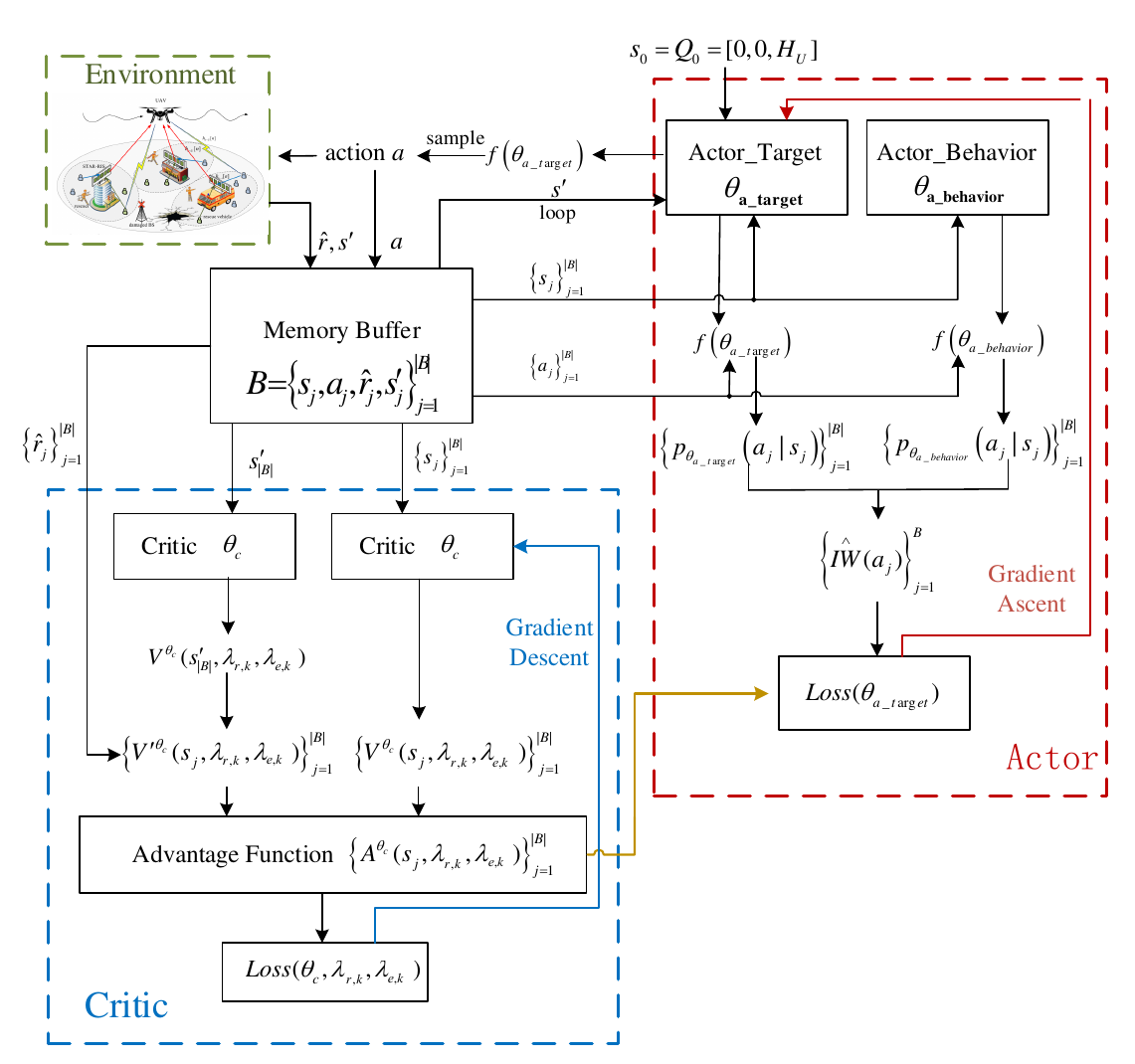}\\
  \caption{Diagram of the Reward Constrained Proximal Policy optimization}\label{Fig3}
  \vspace{-0.6cm}
\end{figure}

\begin{remark}\label{remark2}
Unlike A3C (updated per step) or Policy Gradient (updated per episode), the network parameters ${\theta _c}$ , ${\theta _{a\_target}}$ of the PPO are updated per $|B|$-steps. Besides, the Actor training is guided by the output of the Critic, so ${\eta _c} > {\eta _a}$ is often required to ensure that the Critic update is performed on a faster timescale than that of the Actor.
\end{remark}
The detailed diagram of the reward constrained proximal policy optimization is shown in Fig. \ref{Fig3}.

\subsection{The outer layer : Gradient Descent}
In this subsection, with policy $\pi_{\theta}$ obtained by the reward constrained proximal policy optimization algorithm, we aim to update the Lagrange multipliers ${\lambda _{r,k}}$, ${\lambda _{e,k}}$ in the outer layer of (\ref{P3}) by gradient descent.
The partial derivatives ${\nabla _{{\lambda _{r,k}}}}L({\lambda _{r,k}},{\lambda _{e,k}},\pi_{\theta} )$ and ${\nabla _{{\lambda _{e,k}}}}L({\lambda _{r,k}},{\lambda _{e,k}},\pi_{\theta} )$ are easily derived from (\ref{Lagrange_function}), which are denoted as $ {\nabla _{{\lambda _{r,k}}}}L({\lambda _{r,k}},{\lambda _{e,k}},{\pi_{\theta}} ){\rm{ = }} -  J_{{C_{r,k}}}^{\pi_{\theta}}$ and ${\nabla _{{\lambda _{e,k}}}}L({\lambda _{r,k}},{\lambda _{e,k}},{\pi_{\theta}} ){\rm{ = }} - J_{{C_{e,k}}}^{\pi_{\theta}}$ respectively. Then, the Lagrange multipliers are updated as
\begin{equation}\label{lamda_update1}
{\lambda _{r,k}}^\prime  = {\Gamma _\lambda }\left[ {{\lambda _{r,k}} - {\eta _{l,r}}{\nabla _{{\lambda _{r,k}}}}L({\lambda _{r,k}},{\lambda _{e,k}},{\pi_{\theta}} )} \right] = {\Gamma _\lambda }\left[ {{\lambda _{r,k}} + {\eta _{l,r}}J_{{C_{r,k}}}^{{\pi_{\theta}}}} \right],
\end{equation}
\begin{equation}\label{lamda_update2}
{\lambda _{e,k}}^\prime  = {\Gamma _\lambda }\left[ {{\lambda _{e,k}} - {\eta _{l,e}}{\nabla _{{\lambda _{e,k}}}}L({\lambda _{r,k}},{\lambda _{e,k}},{\pi_{\theta}} )} \right] = {\Gamma _\lambda }\left[ {{\lambda _{e,k}} + {\eta _{l,e}}J_{{C_{e,k}}}^{{\pi_{\theta}}}} \right],
\end{equation}
where ${\Gamma _\lambda }$ is an operator keeping the Lagrange multipliers positive, that is ${\lambda _{r,k}},{\lambda _{e,k}} \in [0,\infty ]$. ${\eta _{l,r}}$ and ${\eta _{l,e}}$ are the updating step-sizes for ${\lambda _{r,k}}$ and ${\lambda _{e,k}}$ respectively.
\begin{remark}\label{remark3}
As $ J_{{C_r}_{,k}}^\pi$ and $J_{{C_e}_{,k}}^\pi$ are long-term functions, all information over the $N$ time slots is required for (\ref{lamda_update1}) and (\ref{lamda_update2}). In other words, the Lagrange multipliers are updated per episode ($N$-steps). The updating rate should be slower than the learning rates of Critic network and Actor network, that is, ${\eta _c} > {\eta _a} > {\eta _{l,r}}/{\eta _{l,e}}$.
\end{remark}
\subsection{Discussion on the Proposed LRCPPO Algorithm}
We propose the LRCPPO based joint optimization algorithm to find the suboptimal solution of (\ref{P1}), and the procedure is given in Algorithm \ref{LRCPPO}.The agent in the proposed algorithm is acted by a central controller, which could be realized by the Software Defined Networking (SDN). The central controller is able to obtain the data required for training, including the positions of UEs and STAR-RISs and channel information. Also, the optimized policy is transmitted to the UAV, UEs and STAR-RISs by the central controller.

{\bfseries ~Convergence:} The LRCPPO is a combination of the Lagrange relaxation and the PPO, in which a three-timescale approach is involved. On the fast timescale, the penalized reward is estimated by the Critic network; on the intermediate timescale, the parameterized policy is learned by the Actor network; and on the slow timescale, the Lagrange multipliers are updated by the gradient descent. In order to ensure the above three timescales, ${\eta _c} > {\eta _a} > {\eta _{l,r}}/{\eta _{l,e}}$ is required. Based on the Theorem 2 given in \cite{RCPO}, it can be confirmed that as long as ${\eta _c} > {\eta _a} > {\eta _{l,r}}/{\eta _{l,e}}$ is satisfied, the LRCPPO algorithm can converge to a feasible solution. For a full proof to convergency for the three-timescale approximation process, please refer to the Appendix E in \cite{RCPO}.

{\bfseries ~Complexity:}~ Recall that the state dimension of the CMDP is $2$ and the action dimension is $2+K+I+2MI$. The batch size of sample memory is $|B|$. The LRCPPO is mainly composed of three parts: Critic network, Actor network and Lagrange multipliers. The temporal computational complexity of the Critic network for one episode is
\begin{equation*}
    {{\mathcal{O}}_C}\left( {\left\lceil {\frac{N}{{|B|}}} \right\rceil \left( {\left( {|B| + 1} \right)\left( {{Z_{C0}}{Z_{C1}} + \sum\nolimits_{l = 1}^{{L_C} - 2} {{Z_{Cl}}{Z_{C\left( {l + 1} \right)}} + {Z_{C\left( {{L_C} - 1} \right)}}{Z_{C{L_C}}}} } \right) + 3|B|} \right)} \right),
    \vspace{-0.2cm}
\end{equation*}
where ${L_C}$ is the number of layers of the Critic
neural network, ${Z_{Cl}}$ is the number of corresponding neurons, ${Z_{C0}} = 2$ and ${Z_{C{L_C}}} = 1$. Similarly, the temporal computational complexity of the  Actor network for one episode is
\begin{equation*}
    {{\mathcal{O}}_A}\left( {\left\lceil {\frac{N}{{|B|}}} \right\rceil \left( {2|B|\left( {{Z_{A0}}{Z_{A1}} + \sum\nolimits_{l = 1}^{{L_A} - 2} {{Z_{Al}}{Z_{A\left( {l + 1} \right)}} + {Z_{A\left( {{L_A} - 1} \right)}}{Z_{A{L_A}}}} } \right) + 3|B|} \right)} \right),
    \vspace{-0.2cm}
\end{equation*}
where ${L_A}$ is the number of layers of the Actor
neural network, ${Z_{Al}}$ is the number of corresponding neurons, ${Z_{A0}} = 2$ and ${Z_{A{L_A}}} = 2 + K + I + 2MI$. The temporal computational complexity of the Lagrange multipliers updating for one episode is ${{\mathcal{O}}_L}\left( {2KN} \right)$. We assume that the proposed algorithm converges with $G$ episodes, and then the temporal computational complexity of the LRCPPO is given as ${\mathcal{O}}\left( {G\left( {{{\mathcal{O}}_C} + {{\mathcal{O}}_A} + {{\rm O}_L}} \right)} \right)$.

\begin{algorithm}[!h!]
	\caption{LRCPPO based trajectory design, passive beamforming  and resource allocation algorithm}
	\label{LRCPPO}
\begin{algorithmic}[1]
	\STATE Set number of time slots, UE position, STAR-RIS position,UAV initial position ${\mathbf{q}_0}$, minimum rate ${r_{\min }}$, maximum energy ${E_{\max }}$,
    \STATE Input penalized reward function $\hat r(\cdot)$, penalty ${c_{r,k}}(\cdot)$, ${c_{e,k}}(\cdot)$, constraints ${C_{r,k}}(\cdot)$, ${C_{e,k}}(\cdot)$; learning rates ${\eta _3} > {\eta _4} > {\eta _1}/{\eta _2}$,
    \STATE  Initialize memory buffer $B = \{\emptyset\}$, neural network parameters ${\theta _c}$, ${\theta _{a\_target}}$, ${\theta _{a\_behavior}}$, Lagrange multipliers ${\lambda _{r,k}},{\lambda _{e,k}} = 0$ ,
	\FOR{$episode = 0,1,...G$}
    \STATE Initialize state ${s_0} = {\mathbf{q}_0}$,
	\FOR{$n = 0$ \textbf{to} $n = N - 1$}
    \STATE Sample action ${a_n\sim{\pi^{\theta _{a\_target}}}}$, observe next state ${s_{n + 1}}$, penalized reward $\hat r({s_n},{a_n})$ and penalty ${c_{r,k}}({s_n},{a_n})$, ${c_{e,k}}({s_n},{a_n})$,
    \STATE    Add experiences to memory buffer, $B \leftarrow ({s_n},{a_n},{\hat r_n},{s_{n + 1}}) \cup B$,
    \IF {$n\% |B| = {\kern 1pt} {\kern 1pt} 0{\kern 1pt} {\kern 1pt} {\kern 1pt}$ or ${\kern 1pt} {\kern 1pt} {\kern 1pt} n = N - 1$}
    \STATE   ${V'^{{\theta _c}}}({s_{|B| + 1}},{\lambda _{r,k}},{\lambda _{e,k}}) = {V^{{\theta _c}}}({s_{|B| + 1}},{\lambda _{r,k}},{\lambda _{e,k}})$,
    \STATE    ${V'^{{\theta _c}}}({s_j},{\lambda _{r,k}},{\lambda _{e,k}}) = \hat r({\lambda _{r,k}},{\lambda _{e,k}},{s_j},{a_j}) + \gamma {V'^{{\theta _c}}}({s_{j + 1}},{a_{j + 1}},{\lambda _{r,k}},{\lambda _{e,k}})|s$,
    \STATE    \textbf{Critic update:}  ${\theta _c}^\prime  = {\theta _c} - {\eta _3}{\nabla _{{\theta _c}}}Loss({\theta _c})$,
    \STATE    Importance weight: $IW({a_j}) = {p_{{\theta _{a\_t\arg et}}}}\left( {{a_j}|{s_j}} \right)/{p_{{\theta _{a\_behavior}}}}\left( {{a_j}|{s_j}} \right)$,
    \STATE    Advantage function: ${A^{{\theta _c}}}({s_j},{\lambda _{r,k}},{\lambda _{e,k}}) = {V'^{{\theta _c}}}({s_j},{\lambda _{r,k}},{\lambda _{e,k}}) - {V^{{\theta _c}}}({s_j},{\lambda _{r,k}},{\lambda _{e,k}})$,
    \STATE    \textbf{Actor update:} ${\theta '_{a\_t\arg et}} = {\theta _{a\_t\arg et}} + {\eta _4}{\nabla _{{\theta _{a\_t\arg et}}}}Loss({\theta _{a\_t\arg et}})$,
    \STATE    $B = \{ \emptyset \}$,
    \STATE    Update ${\theta _{a\_behaivior}} = {\theta _{a\_target}}$  after c iterations,
    \ENDIF
    \ENDFOR
    \STATE \textbf{Lagrange multipliers update:} ${\lambda _{r,k}}^\prime={\Gamma _\lambda }\left[ {{\lambda _{r,k}} + {\eta _1}J_{{C_{r,k}}}^{{\pi _\theta }}} \right]$, ${\lambda _{e,k}}^\prime={\Gamma _\lambda }\left[ {{\lambda _{e,k}} + {\eta _2}J_{{C_{e,k}}}^{{\pi _\theta }})} \right]$,
    \ENDFOR
    \STATE \textbf{Return} policy parameters ${\theta _{a\_t\arg et}}$
\end{algorithmic}
\end{algorithm}

\section{Simulation Results}
In this section, numerical results are provided to verify the performance of the LRCPPO based joint optimization algorithm and the proposed STAR-RIS assisted UAV NOMA network. In the simulation, we consider a disaster area with the size of $800$ m $\times$ $800$ m. The number of STAR-RISs is $3$ and their positions are fixed at $[400,600,10]$, $[200,300,10]$ and $[600,200,10]$. In addition, the element spacing of the STAR-RIS is assumed as half of the carrier wavelength. UEs are distributed randomly in the area. The UAV's initial location is set to ${\mathbf{q}_0}= [0,0,200]$. The main simulation setups are shown in  Table \ref{table2}.

Three cases with different constraints are considered in the simulation. In case~1,  there are no constraints on UE rate or UE energy. In case~2, the minimal average rate requirement is set as $300000$ bps and the maximal available energy for UE is $180$ J, that is, $r_{\min } = 300000$  bps and $E_{\max } = 180$ J. In case~3, the constraints are more stringent, where the lower limit for UE rate and the upper limit for UE energy are $1000000$ bps and $90$ J, respectively.

\begin{table}[!h]
	\caption{ Simulation Parameters}
    \label{table2}
	\begin{center}
		\begin{tabular}{|p{7cm}<{\centering}|c|}
            \hline
            \textbf{Parameter} & \textbf{Value}  \\       \hline
            Number of UEs          & $K = 12 $ \\        \hline
			Length of one time slot          & $\tau= 1$ s \\        \hline
            Duration                         & $T = 200$ s \\        \hline
            Available bandwidth            & $BW = 1$ MHz \\ \hline
            Maximal UE transmission power   & ${P_{\max}} = 1$ W \\ \hline
            UAV fight height                 & ${H_U} = 200$ m \\ \hline
            Maximal UAV fight speed      & ${v_{\max}} = 10$ m/s \\ \hline
            Carrier frequency               & ${f_c} = 3.6$ GHz \\ \hline
            Noise power                     & ${\sigma ^2} = -60$ dBm  \\ \hline
            Rician factor                   & $G_{i - U} (G_{i - k},G_{k - U}) = 10$ dB \\ \hline
            Channel gain at 1 meter         &${\xi _0} = -45$ dB    \\ \hline
		\end{tabular}
	\end{center}
\end{table}
\vspace{-1.2cm}
\begin{figure}[htbp]
  \centering
  \includegraphics[width=4.5in]{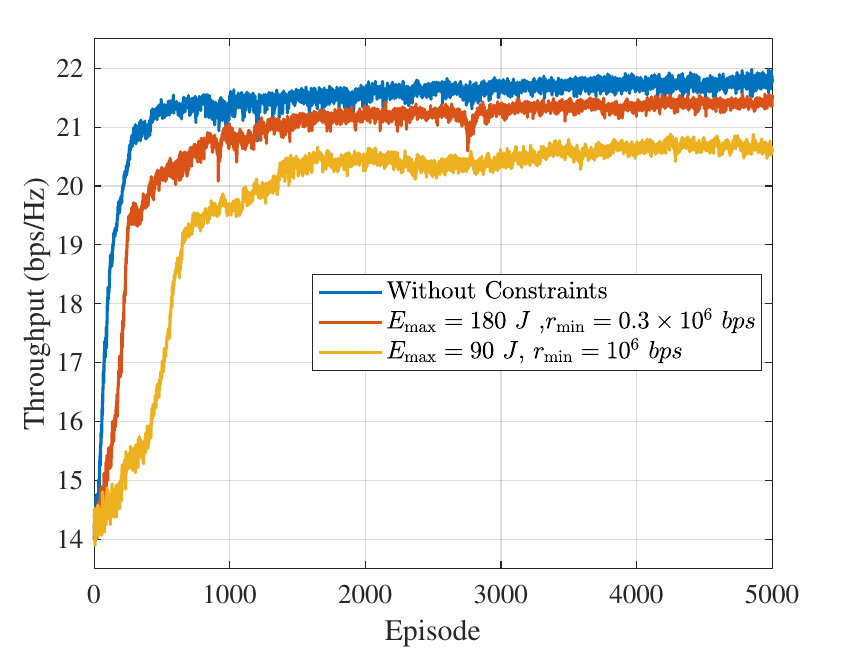}\\
  \caption{Convergence of the proposed algorithm with different constraints.}\label{Throughput}
  \vspace{-0.2cm}
\end{figure}

\begin{figure}[htbp]
  \vspace{-0.2cm}
  \centering
  \includegraphics[width=4.5in]{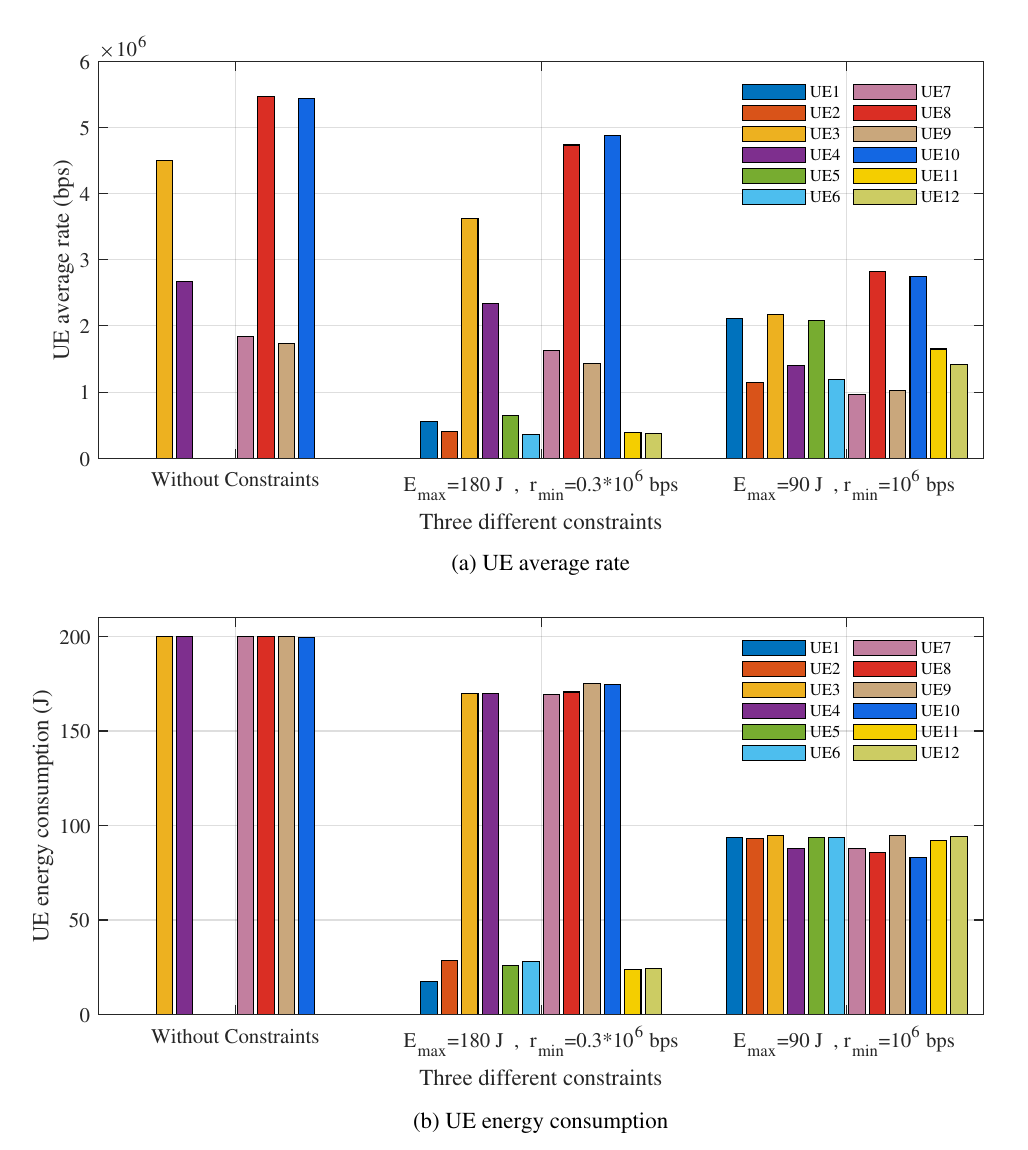}\\
  \vspace{-0.2cm}
  \caption{UE average rate and UE long-term energy consumption with different constraints.}\label{UE}
  \vspace{-0.2cm}
\end{figure}

Figs. \ref{Throughput}--\ref{UE} show the feasibility of the proposed LRCPPO based joint optimization algorithm in the above three cases. The convergence is verified in Fig. \ref{Throughput}, where the number of STAR-RIS elements is $40$. Fig. \ref{Throughput} shows that as the training progresses, the uplink throughput in all three cases increases quickly and finally tends to be stable. It is worth noting that the convergence speed of the proposed algorithm slows down with stricter constraints. This is because more numbers of iterations are needed for updating the Lagrange multipliers to satisfy stricter constraints, and the parameters of the Actor network and the Critic network are updated accordingly. In addition, the optimized throughput for case~1, case~2 and case~3 decrease successively, which are about $21.89 $ bps/Hz, $21.37$ bps/Hz, and $20.76$ bps/Hz. This shows that to satisfy the UE rate constraints and UE energy consumption constraints, the overall performance of the system is sacrificed.

Fig. \ref{UE} shows the average rates and the cumulative energy consumption of $12$ UEs in three cases. It can be observed that the rate constraints (\ref{a}) and the energy consumption constraints (\ref{b}) and (\ref{c}) of $12$ UEs are all well satisfied by the proposed LRCPPO algorithm for both case~2 and case~3. The average rates without constraints vary greatly among $12$ UEs, where the maximum rate is able to reach $5.5$ Mbps while the minimum rate is close to $0$. The variance is narrowed in case~2 and significantly smaller in case~3. The cumulative energy consumption of UEs in three cases possesses the same performance as the average rates.
The above phenomenon shows that stricter constraints lead to stronger UE fairness, which is of great importance in emergency communication scenarios.
In addition, Fig. \ref{UE} illustrates that the constraint on the minimum UE average rate can increase the rates of some UEs, while the achievable rates of other UEs are reduced due to the limited UE energy. The increased component is smaller than the decreased component, so the overall throughput of the network decreases with stricter constraints, which provides further explanations for the results in Fig. \ref{Throughput}.

\begin{figure}[htbp]
  \centering
  \includegraphics[width=4.5in]{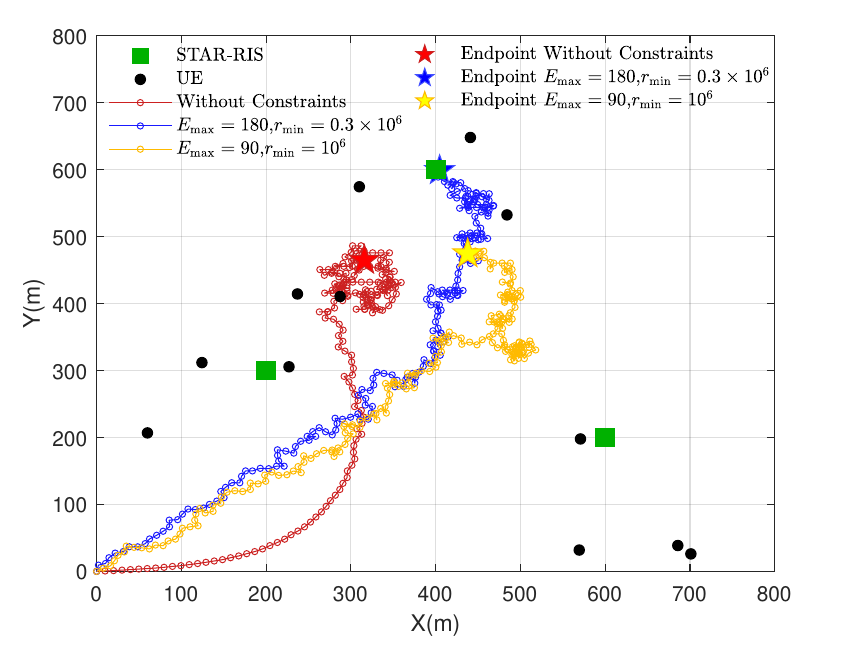}\\
  \caption{Optimized UAV trajectory with different constraints.}\label{UAV_PATH}
  \vspace{-0.2cm}
\end{figure}

In Fig. \ref{UAV_PATH}, the optimized UAV trajectories in three cases are plotted. In the duration $T$, the UAV takes off at the same point but ends up in different positions in three cases. Without constraints, the UAV passes through UEs and then hovers near a globally superior position. The globally optimal position is jointly determined by the path loss of both direct link and STAR-RIS aided link for $12$ UEs. In case~2 and case~3, the UAV adaptively adjusts its trajectory and enlarges its active area, so as to take into account each UE and contribute to satisfying the constraints.

Then we evaluate the performance of the proposed LRCPPO based joint optimization algorithm in comparison to the following benchmark algorithms:
\begin{enumerate}
  \item \emph{Reward shaping based PPO}: As mentioned in Remark \ref{remark1}, reward shaping is a common method for solving constrained long-term optimization problems with reinforcement learning \cite{uav_ris,reward_shaping2,reward_shaping3}. For our proposed problem, the reward function after reward shaping is defined as
     \begin{equation*}
      r({s_n},{a_n}) = \sum\limits_{k = 1}^K {{r_k}[n]}  - {\chi _r}\sum\limits_{k = 1}^K {{c_{r,k}}\left( {{s_n},{a_n}} \right)}  - {\chi _e}\sum\limits_{k = 1}^K {{c_{e,k}}\left( {{s_n},{a_n}} \right)},
      \end{equation*}
      where ${\chi _r}$ and ${\chi _e}$ are coefficients used to balance the values. To be fair, we also adopt the PPO algorithm in the reward shaping.
  \item  \emph{Zero phase shift}: The phase shifts of the STAR-RIS elements are all fixed at zero, while other variables are still determined by the LRCPPO.
  \item  \emph{Random phase shift}: The phase shifts of the STAR-RIS elements are randomly generated, while other variables are still determined by the LRCPPO.
  \item  \emph{Equal time allocation}: The time slot is equally allocated to the left UEs and the right UEs, that is  $\tau _i^L[n] = \tau _i^R[n]=\frac{\tau }{2}$. The other variables are still determined by the LRCPPO.
  \item  \emph{Circular UAV trajectory}: The UAV flies at the maximum speed ${v_{\max }}$ along a circular path with $[400,400,200]$ as the center and a radius of $400$ m. The other variables are still determined by the LRCPPO.
\end{enumerate}

\begin{figure}[!h]
  \centering
  \vspace{-0.3cm}
  \includegraphics[width=3.8in]{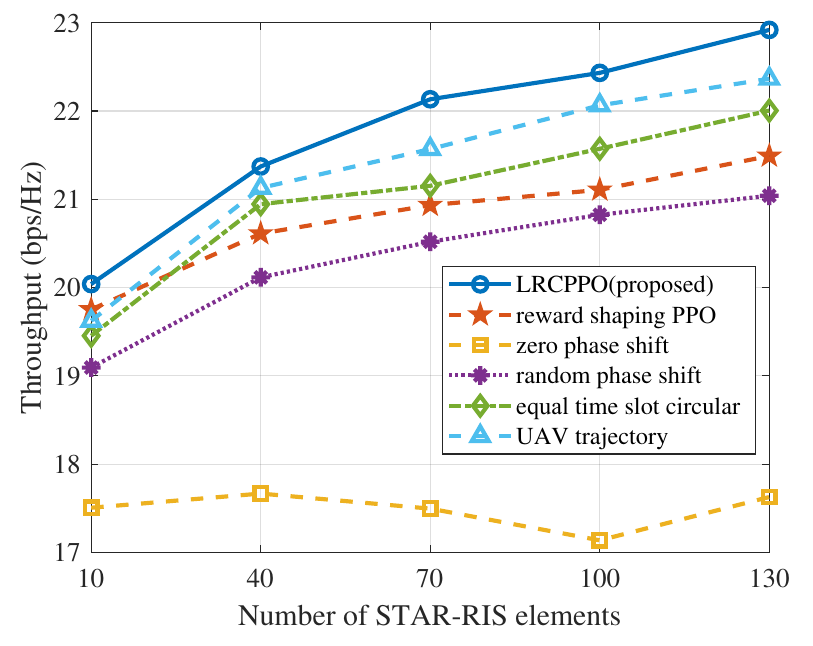}\\
  \vspace{-0.2cm}
  \caption{Throughput under different algorithms versus the number of STAR-RIS elements with $r_{\min } = 300000$ bps $\&~ E_{\max } = 180$ J.}\label{alg1}
\end{figure}

\begin{figure}[!h]
  \vspace{-0.3cm}
  \centering
  \includegraphics[width=3.8in]{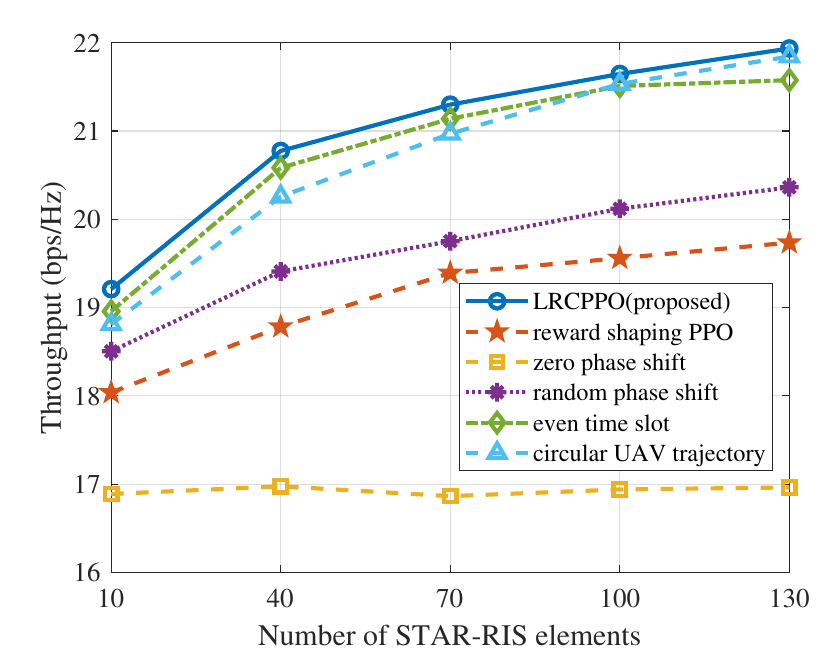}\\
  \vspace{-0.1cm}
  \caption{Throughput under different algorithms versus the number of STAR-RIS elements with $r_{\min } = 1000000$ bps $\&~ E_{\max } = 90$ J.}\label{alg2}
\end{figure}

Fig. \ref{alg1} and Fig. \ref{alg2} illustrate the performance of the proposed algorithm in case~2 and case~3. Firstly, it is seen that the throughput obtained by the proposed LRCPPO, reward shaping based PPO, random phase shift, equal time allocation, and circular UAV trajectory all increases with the number of STAR-RIS elements in Fig. \ref{alg1} and Fig. \ref{alg2}. However, the throughput of the zero phase shift appears to be independent of the number of STAR-RIS elements and performs much worse than that of the other five algorithms. This shows that the phase shifts of the STAR-RIS have to be reasonably designed, so as to achieve the performance gain.
In addition, no matter in case 1 or case 2, the uplink throughput achieved by the proposed LRCPPO algorithm is significantly greater than that obtained by the reward shaping based PPO algorithm. This verifies that, as explained in Remark \ref{remark1}, due to manual selection of the penalty coefficients, the non-adaptive reward shaping PPO algorithm inevitably sacrifices the overall network performance to satisfy the constraints. Differently, the penalty coefficients in the LRCPPO algorithm are acted by the Lagrange multipliers, and updated iteratively by gradient descent. Therefore, the LRCPPO is able to make up for the vital shortcoming of the reward shaping, and bring evident improvement in network performance.
In fact, in both case~2 and case~3, the throughput of the proposed algorithm is the largest among all six algorithms, which verifies the superiority of the proposed algorithm. Furthermore, comparing Fig. \ref{alg1} and Fig. \ref{alg2}, we can find that this superiority is more pronounced with $r_{\min } = 300000$ bps $ \&~ E_{\max } = 180$ J than $r_{\min } = 1000000$ bps $\&~ E_{\max } = 90$ J. This is because that as mentioned earlier, the stricter constraints on UE rate and UE energy significantly limit the maximum achievable network throughput.


\begin{figure}[!h]

  \centering
  \includegraphics[width=3.8in]{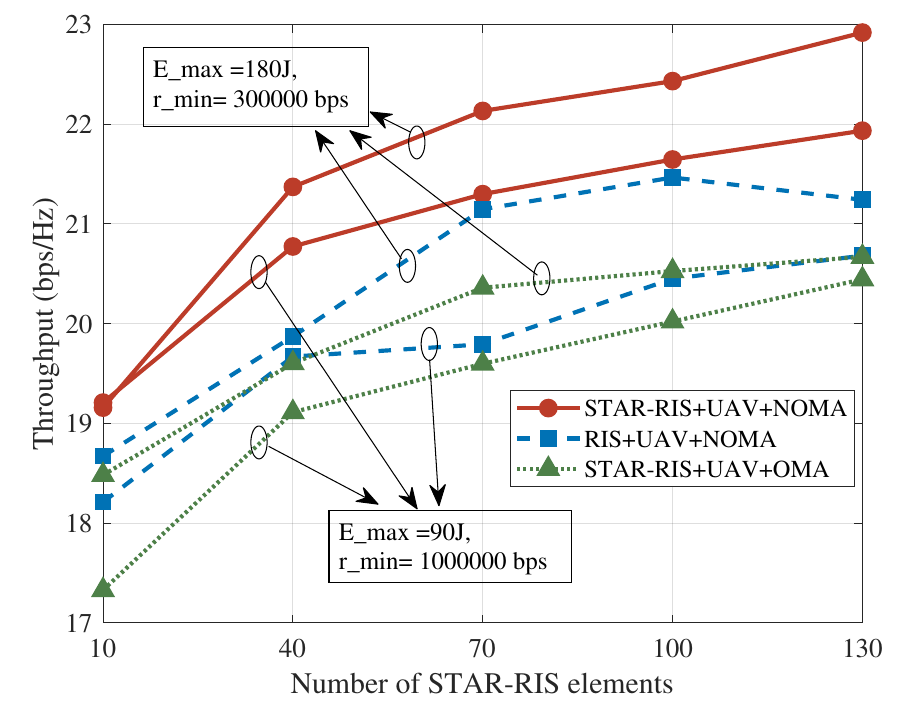}\\
  \vspace{-0.1cm}
  \caption{Throughput under different network framework versus the number of STAR-RIS elements.}\label{framework}
  \vspace{-0.3cm}
\end{figure}

\newpage
In Fig. \ref{framework}, we demonstrate the performance of the proposed STAR-RIS assisted UAV NOMA communication network, by comparing it with two benchmark networks. One of the benchmarks is reflecting-only RIS assisted UAV NOMA communication network, where the RIS is only able to serve the UEs located at the same side of the RIS as the UAV.  The other is STAR-RIS assisted UAV OMA communication network, where the UEs at the same side of the same STAR-RIS, ${\mathcal{K}}_i^L$ (${\mathcal{K}}_i^R$), perform uplink data transmission by orthogonal frequency division multiple access (OFDMA). All lines in the figure are obtained by the proposed LRCPPO based algorithm. First of all, it is observed that with the increase in the number of (STAR-)RIS elements, the throughput of three different networks all shows an obvious upward trend. Secondly, no matter in case~2 ($r_{\min } = 300000$ bps $ \&~ E_{\max } = 180$ J) or case~3 ($r_{\min } = 1000000$ bps $ \&~ E_{\max } = 90$ J), with the same number of elements, the throughput of the proposed network is higher than that of the reflecting-only RIS assisted network, and much higher than that of the OMA-based network. This reveals that the combination of STAR-RIS and NOMA makes important contributions to the improvement of network performance. Besides, for the same network architecture, the throughput in case~2 is always higher than that in case~3, which illustrates again that the overall performance is sacrificed for meeting the individual UE constraints.

\section{Conclusion}

In this paper, we investigated a novel STAR-RIS assisted UAV NOMA emergency communication network, where multiple STAR-RISs were employed to assist the UAV-mounted BS and NOMA was involved for further improvement. We formulated a long-term uplink communication throughput maximization problem subject to the constraints on minimum UE average rate and maximum UE available energy in post-disaster emergency communication scenarios. A Lagrange based reward constrained proximal policy optimization algorithm was developed. Numerical results revealed that the constraints were satisfied well and the UAV trajectory can be adaptively adjusted by the proposed LRCPPO algorithm. The performance of the proposed algorithm was verified to be the best among several benchmark algorithms. It also showed that the proposed STAR-RIS assisted UAV NOMA framework can significantly improve the performance over the benchmark reflecting-only RIS and OMA schemes. The joint optimization of STAR-RIS deployment and UAV trajectory will be investigated in our future study.

\bibliographystyle{IEEEtran}
\renewcommand{\baselinestretch}{1.14}
\bibliography{mybib}

%

\end{document}